\documentclass[5p]{elsarticle}

\usepackage[utf8]{inputenc}
\usepackage{graphicx}
\usepackage{xcolor}
\definecolor{refC}{rgb}{0,0,0.75}
\usepackage[unicode,colorlinks,linkcolor=refC]{hyperref}

\usepackage{import}
\usepackage{multirow}
\usepackage{makecell}
\usepackage{amsmath}

\usepackage{booktabs}

\usepackage{listings}
\usepackage{float}
\usepackage{stfloats}
\usepackage{placeins}

\usepackage{soul}
\soulregister\cite7
\soulregister\ref7
\soulregister\pageref7
\soulregister\autoref7

\newcommand*{\fullref}[1]{\hyperref[{#1}]{\autoref*{#1}: \nameref*{#1}}}
\newcommand*{\shortref}[1]{\hyperref[{#1}]{\ref*{#1} \nameref*{#1}}}

\colorlet{punct}{red!60!black}
\definecolor{background}{HTML}{EEEEEE}
\definecolor{delim}{RGB}{20,105,176}
\colorlet{numb}{magenta!60!black}

\lstdefinelanguage{json}{
    basicstyle=\normalfont\footnotesize\ttfamily,
    numbers=none,
    numberstyle=\scriptsize,
    stepnumber=1,
    numbersep=8pt,
    columns=flexible,
    keepspaces=true,
    showspaces=false,
    breaklines=true,
    frame=lines,
    backgroundcolor=\color{background},
    literate=
     *{0}{{{\color{numb}0}}}{1}
      {1}{{{\color{numb}1}}}{1}
      {2}{{{\color{numb}2}}}{1}
      {3}{{{\color{numb}3}}}{1}
      {4}{{{\color{numb}4}}}{1}
      {5}{{{\color{numb}5}}}{1}
      {6}{{{\color{numb}6}}}{1}
      {7}{{{\color{numb}7}}}{1}
      {8}{{{\color{numb}8}}}{1}
      {9}{{{\color{numb}9}}}{1}
      {e}{{{\color{numb}e}}}{1}
      {:}{{{\color{punct}{:}}}}{1}
      {,}{{{\color{punct}{,}}}}{1}
      {\{}{{{\color{delim}{\{}}}}{1}
      {\}}{{{\color{delim}{\}}}}}{1}
      {[}{{{\color{delim}{[}}}}{1}
      {]}{{{\color{delim}{]}}}}{1},
}

\journal{Computer Communications}

\begin{document}

\begin{frontmatter}

\title{Flow length and size distributions in campus Internet traffic}

\author{Piotr Jurkiewicz\corref{mycorrespondingauthor}}
\cortext[mycorrespondingauthor]{Corresponding author}
\ead{jurkiew@agh.edu.pl}

\author{Grzegorz Rzym}
\author{Piotr Boryło}

\address{Department of Telecommunications, AGH University of Science and Technology, Kraków, Poland}

\begin{abstract}
The efficiency of flow-based networking mechanisms strongly depends on traffic characteristics and should thus be assessed using accurate flow models. For example, in the case of algorithms based on the distinction between elephant and mice flows, it is extremely important to ensure realistic flows' length and size distributions. Credible models or data are not available in literature. Numerous works contain only plots roughly presenting empirical distribution of selected flow parameters, without providing distribution mixture models or any reusable numerical data.
This paper aims to fill that gap and provide reusable models of flow length and size derived from real traffic traces. Traces were collected at the Internet-facing interface of the university campus network and comprise four billion layer-4 flow (275 TB). These models can be used to assess a variety of flow-oriented solutions under the assumption of realistic conditions.
Additionally, this paper provides a tutorial on constructing network flow models from traffic traces. The proposed methodology is universal and can be applied to traffic traces gathered in any network. We also provide an open source software framework to analyze flow traces and fit general mixture models to them.
\end{abstract}



\end{frontmatter}

\section{Introduction}
\label{introduction}

Flow-based switching and routing has been gaining the attention of researchers for a quite some time \cite{Kreutz:Survey}. It can be advantageous in comparison to per-packet switching, especially with regard to traffic engineering \cite{akyildiz2014roadmap}, quality of service (QoS) \cite{10.1145/2071389.2071394} or security \cite{7568520}. For example, flow routing enables multipath and adaptive approaches, which are impossible to achieve in per-packet routing due to routing loops and route-flapping constraints, respectively \cite{FAMTAR_IMPLEMENTATION}.


The efficiency of numerous flow-based solutions strongly depends on traffic characteristics, and thus, should be assessed based on realistic and accurate flow models. An example of such solutions are traffic engineering mechanisms exploiting the heavy-tailed nature of IP flows. To the best of our knowledge, the first paper exploring such a possibility is \cite{rexford-long-flows}, in which the authors proposed heuristic that differentiates traffic into \emph{elephant} and \emph{mice} flows. Then, assuming that \emph{elephant} flows have a more significant impact on network performance, this type of traffic is routed adaptively to the current network load, while flows classified as \emph{mice} are handled using the shortest paths. Recently, the heavy-tailed nature of IP flows is being exploited to reduce management overheads in software-defined networking (SDN). For example, in work \cite{Mu:SDNFlowEntry}, the authors employed a reinforcement learning approach to detect \emph{elephant} flows in advance to limit the number of flow entries in forwarding tables.

To reliably evaluate such ideas, realistic distributions of flows' length and size must be ensured. Unfortunately, such a data is not available in the literature. For example, the authors of \cite{rexford-long-flows} used distributions extracted from their own traffic measurements, but they did not provide any reusable data and their trace was limited to only one week. By contrast, the authors of \cite{Mu:SDNFlowEntry} assumed a 1:9 constant ratio between \emph{elephant} and \emph{mice} layer-4 flows and fixed flow sizes of 25.6 MB and 256 KB, respectively. Such assumptions are not only arbitrary, they also often do not correspond to reality, as we show in this work.

The lack of realistic models negatively impacts on the credibility of results presented in numerous papers. Moreover, different and arbitrary assumptions in various works exclude the possibility to effectively compare different solutions. As we show in related works, all the papers attempting address this issue provide plots presenting empirical probability density functions (PDFs) or cumulative distribution functions (CDFs) of selected flow parameters at best. None of these papers provide distribution mixture models or even reusable numerical data of any kind.

\begin{table*}[!h]
\scriptsize
\caption{Examples of research areas where the proposed model can be used}
\begin{center}
\begin{tabular}{@{}p{6cm} p{10cm} @{}}
\toprule
\textbf{Research area} & \textbf{Usage example} \\
\midrule
AI-based networking & Learning and evaluation of novel AI-based routing algorithms \\
\midrule
Big Data driven networking & Generating large amounts of flow records for analysis \\
\midrule
Traffic engineering & Evaluation of flow-based traffic engineering methods \\
\midrule
SDN management & Selecting elephant detection thresholds for reduction of flow table size requirements \\
\midrule
Algorithms \& protocols performance evaluation & Assessment of novel algorithms and protocols in realistic conditions \\
\midrule
Equipment performance evaluation & Performance evaluation of new hardware/chips at realistic flow rate \\
\bottomrule
\end{tabular}
\label{tab-applications}
\end{center}
\end{table*}

This paper's goal is to provide accurate flow statistics and reusable distribution mixture models of flow's length and size for any researchers who may need such a data. We believe that these models can be considered as general models of typical Internet traffic, and thus, widely used in numerous applications, including AI and Big Data. Examples of such the applications are summarized in~\autoref{tab-applications}. Furthermore, we provide a tutorial and software for building similar models based on data gathered in any network.

The structure of this paper is as follows. First, we present the methodology with a tutorial covering the following steps:

\begin{itemize}

\item collecting flow records,

\item cleaning the data,

\item merging split flow records,

\item data binning and plotting,

\item fitting mixture models to data,

\item generating realistic traffic based on these models.

\end{itemize}

For each step, we provide tips and highlight caveats and possible pitfalls. In addition to the methodology and tutorial, we provide an open source software framework comprising tools aimed at performing these steps. The framework is designed with big data analysis capabilities in mind. Specifically, it supports out-of-core computing, making possible to analyze data which exceeds available memory. Moreover, most processing steps can be scaled horizontally using the well-established map-reduce technique. Therefore, provided implementation is not limited in terms of the number of processed flow records. Together, the provided methodology and framework create an opportunity for any interested parties to extract traffic characteristics from their networks and validate any potential mechanisms before applying them in the production environment.

Then, we use the framework to apply the methodology to the real traffic traces in order to extract models of flow length and size. Traces cover a thirty-day period of layer-4 flows (four billion flows, defined by 5-tuple, 275 TB of transmitted data) and were collected on the Internet-facing interface of a large wired university network. This is several orders of magnitude more than in previous analyses which mostly comprised tens of millions of flows. Flows number, and total sum of packet and octet distributions are extracted, analyzed and modeled as functions of both flow length (in packets) and flow size (in bytes). In previous works, only selected distributions were presented, without any models or reusable numerical parameters (see \fullref{related-works}).

Finally, along with the framework source code, we make the data publicly available. This makes our results reusable and fully reproducible, increasing the value of the tutorial part of this work:

\smallskip

\noindent \url{https://github.com/piotrjurkiewicz/flow-models}

\section{Related works}
\label{related-works}

\begin{table*}[!h]
\scriptsize
\caption{Distribution plots presented in related works}
\begin{center}
\begin{tabular}{@{}l c@{}}
\toprule
Distribution of \textbf{flows} in function of flow's \textbf{length}: & \cite{rexford-long-flows} \cite{Pustisek:Emperical} \cite{kim2006characteristic} \cite{Ryu01internetflow} \cite{Fang:Seamless} \cite{Guan:Dynamic} \cite{Qian:flow} \\
\midrule
Distribution of \textbf{packets} in function of flow's \textbf{length}: & \cite{rexford-long-flows} \\
\midrule
Distribution of \textbf{flows} in function of flow's \textbf{size}: & \cite{Pustisek:Emperical} \cite{kim2006characteristic} \cite{Ryu01internetflow} \cite{Fang:Seamless} \cite{zhang2002characteristics} \cite{Qian:TCPRevised} \cite{Brownlee-understanding} \cite{Papagiannakit:Impact} \cite{Benson:Network} \cite{Lan03onthe} \cite{LAN200646} \cite{megyesi2013analysis} \cite{antunes2016estimation} \\
\midrule
Distribution of \textbf{octets} in function of flow's \textbf{size}: & \cite{megyesi2013analysis} \\
\bottomrule
\end{tabular}
\label{tab-related-distributions}
\end{center}
\end{table*}

\begin{table*}[!h]
\scriptsize
\caption{Comparison with most significant related works}
\begin{center}
\begin{tabular}{@{}l p{3.5cm} p{3.5cm} p{3.5cm} p{3.5cm} @{}}
\toprule
\textbf{Work/Property} & \cite{rexford-long-flows} & \cite{Pustisek:Emperical} & \cite{megyesi2013analysis} & This work\\
\midrule
Collection point & Access point of the AT\&T WorldNet network & University of Ljubljana campus WAN port & CAIDA traces and Budapest University (BME) campus WAN port & AGH University of Science and Technology WAN port \\
\midrule
Date & June 1997 & June 2007 & December 2012 & June 2015 \\
\midrule
Period & 1 week & 24 hours & 4 minutes (CAIDA) and 6 minutes (BME) & 30 days\\
\midrule
Number of flows & 795 thousand & 50 million & 4 million (CAIDA) and 264 thousand (BME) & 4 billion\\
\midrule
Main contribution & Proportion of packets in long flows & Fitted (single) distributions & CDF plots, packet interarrival time analysis & Fitted precise distribution mixtures, reusable CDF numerical data, reproducible\\
\midrule
Crucial assumptions & Inactive timeout equal to 60~s, flows truncated to 1000~s & Inactive timeout unknown, flows truncated to 65535~s & Inactive timeout unknown, flows truncated to 4 and 6 minutes & Inactive timeout equal to 15~s, flows not truncated\\
\midrule
Significant limitations & No models/reusable numerical data & Only selected applications analyzed & CDF presented only graphically & Focused only on flow length and size\\
\bottomrule
\end{tabular}
\label{tab-related-work}
\end{center}
\end{table*}


To our knowledge, no other paper jointly provides tutorial style methodology to extract accurate flow characteristics. Furthermore, we are unaware of either any software framework able to determine such characteristics or any previous work providing reusable flow model reflecting general Internet traffic. Some works provide only selected traffic properties, without trying to fit accurate mixture models. Such works are briefly introduced below.

The contribution of paper \cite{Pustisek:Emperical} is the most similar to our work. The authors also calculate flow statistics based on the traces originating from NetFlow protocol. The output of the performed analyses are empirical CDFs of flow length, size and duration. What distinguishes work \cite{Pustisek:Emperical} is that the authors also fit particular distributions to the data and provide complete descriptive parameters. However, the achieved accuracy is much worse than our work. This is mainly due to the fact that single distributions are considered instead of distribution mixtures utilized in our models. Furthermore, a very important difference is that the authors in \cite{Pustisek:Emperical} examine only selected transport layer ports, representing applications like peer-to-peer (P2P), web and TCP-big. Therefore, the proposed models cannot be used to represent general Internet traffic.

Plots presenting empirical CDFs of flow length, size and duration are also presented in \cite{kim2006characteristic}. However, source data is outdated (originates from traces collected in 2004) and instead of fitting distributions, the authors only provided values of functions at selected subsets of points.

Papers \cite{zhang2002characteristics} and \cite{Qian:TCPRevised} are also widely cited; both provide graphical representations of flow size, duration and rate distributions, but lack numerical data. Contrastingly, in \cite{lee2007passive} the authors focused solely on the flow duration.

Articles \cite{Ryu01internetflow} \cite{Fang:Seamless} \cite{Guan:Dynamic} \cite{Qian:flow} \cite{Brownlee-understanding} \cite{Papagiannakit:Impact} \cite{Benson:Network} \cite{Limmer:Flow} show selected distribution plots, but without providing any data, what makes them hardly useful in further research. The last article, in addition to the CDFs mentioned in the table, provides CDF of active flows, flow interarrival times and packets interarrival times.

In articles \cite{Lan03onthe} and \cite{LAN200646} in addition to the classification according to the table, there are figures where rate vs. size and bandwidth vs. duration presented.

More recent data was used in \cite{megyesi2013analysis}, where the authors analyzed traffic traces collected in 2012 and originating from the CAIDA\footnote{\url{http://www.caida.org}} and Budapest University campus Internet facing port. However, the provided output is limited solely to the graphical presentation of CDFs of overall, flow and packet sizes. Additional contributions concern to some analyses of packet inter-arrival times and considerations about the contribution of \emph{elephant} flows to the overall traffic.

The output of traffic analyses is even more limited in \cite{quan2010characteristics}, where the authors provide only CDFs of flows and total bytes in function of flow duration. However, the paper is worth mentioning due to its valuable contribution regarding methodology aimed at merging flow records based on packet headers. Finally, no distributions are presented in \cite{Estan-newDirections}, where authors focused on other important aspects. They provided a plot presenting the contribution of different flows to the overall traffic. Such analyses can be especially useful to evaluate mechanisms based on the distinction between \emph{mice} and \emph{elephant} flows. Unfortunately, the work does not provide any reusable numerical data.

Finally, there is a large group of work on traffic distributions of a single network service (e.g. HTTP, video streaming, or voice over IP (VoIP)): \cite{hernandez2004variable} \cite{ramaswami2014modeling} \cite{downey2005lognormal} \cite{garsva2015packet} \cite{Casilari:Modeling} \cite{Pries:HTTP} \cite{Kuan:Game} \cite{Chang:Traffic} \cite{Farber:Network} \cite{Drago:Dropbox} \cite{Goncalves:Dropbox} \cite{Mah:empirical} \cite{Xiaowei:Designing} \cite{Waldmann:Traffic} \cite{SILVA:Live} \cite{Veloso:hierarchical} \cite{Toral:Accurate} Therefore, these works cannot be considered as universal enough to be representative for the general Internet load.

In addition, there is a series of works that refer to extracting the distribution of flows from packet samples: \cite{antunes2016estimation}, \cite{duffield2003estimating}, \cite{yang2009sample} and \cite{antunes2019regularized}. These papers are orthogonal to our work as their main focus is on the estimation of distributions from sampled traffic (incomplete data), which was out of consideration of this work.

Examples of works that take advantage of heavy-tailed nature of IP flows in SDN are \cite{curtis2011devoflow}, \cite{7810727} or \cite{8385221}. DevoFlow \cite{curtis2011devoflow} is a complete TE system based on modified OpenFlow switches, which key feature is the reduction of OpenFlow overhead by focusing on significant flows. The paper \cite{7810727} proposes the ZOOM algorithm based on packet counters for lightweight elephant detection in SDN networks for efficient flow based traffic management. On the other hand, the authors of \cite{8385221} model the problem of flow table occupancy reduction by focusing on elephant flows as the knapsack problem.

None of the works provide parameters of distribution mixture models fitted to the network traffic that can be considered as an approximation of the general Internet load. Furthermore, only selected works provide numerical values of distributions at selected points, while most of them are limited to graphical presentation of CDFs plots. Moreover, none of these papers provide any software. We summarize presented distributions by paper in \autoref{tab-related-distributions}. We also provide a comparison of our paper with the most prominent other works in \autoref{tab-related-work}.

\section{Methodology}

This section covers steps aimed at collecting and analyzing flow traces from the network, as well as constructing flow models that accurately describe the traffic. For each stage, numerous tips and possible pitfalls are provided to reveal all the lessons learned during the research.

\begin{figure}[!htb]
\centering
\includegraphics[scale=0.80]{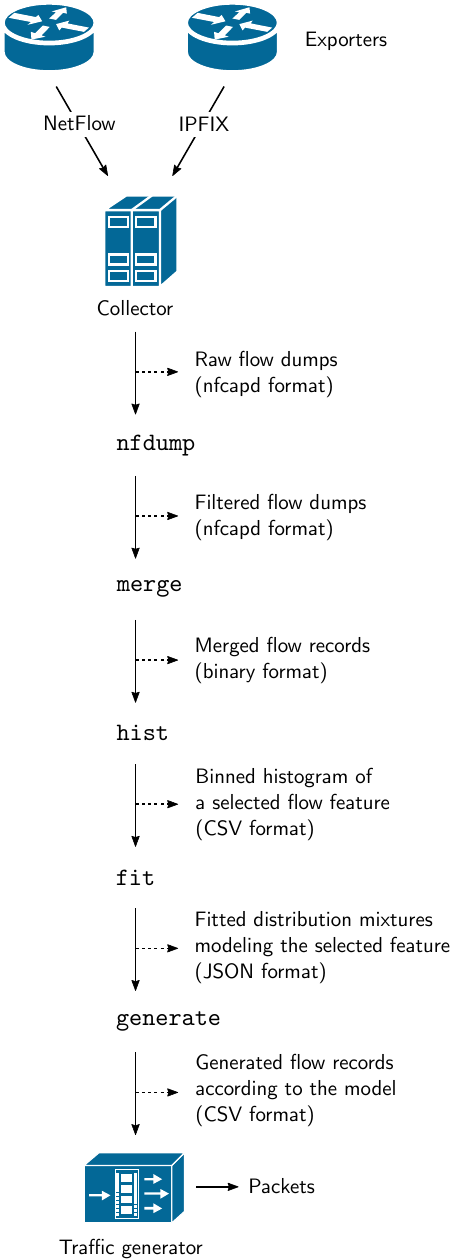}
\caption{The schema of data processing pipeline.}
\label{schema}
\end{figure}

The overall data pipeline is as follows. First, all flow records have to be collected. Next, before any further processing, the data need to be cleaned and filtered. Since long lasting flows may be reported multiple times due to triggering procedures in the exporters, such flow records have to be found and merged back. The next step is reduction of data passed to the modeling by binning it. Fitting of a general mixture model, approximating the collected data, follows afterwards. Fitted model can be used to mimic real traffic in simulators or traffic generators. Schema of the whole pipeline is presented on \autoref{schema}.

\subsection{Flow definition}

The most universal definition of traffic flow in packet network is a sequence of packets which share a common property. Depending on the purpose, flows can be defined at various layers and levels of granularity: layer 4 (five-tuple) or layer 3 (source-destination address pair or destination address/prefix). Any network operator have to decide which fields to consider and this decision usually depends on the operator's scope of operation and traffic engineering purposes. One extreme example are Tier 1 operators which may operate on autonomous system level without even considering IP addresses. However, the most common approach is to determine flow properties based on the packet header fields and to consider source and destination addresses, not exclusively, in different network layers, e.g. data, network, or transport.

In our research, flow is defined as a unidirectional sequence of packets that share the same five-tuple: IP source address, IP destination address, source port, destination port, and transport layer protocol type. All these fields are placed on fixed positions in IP packets and are directly accessible without using deep-packet inspection. Such an approach follows the NetFlow/IPFIX flow concept~\cite{rfc3917} and enables application-oriented approach traffic analysis. However, as such a flow definition may not be valid for all networks, our contribution can be adjusted to other definitions and operator needs.

\subsection{Collecting of flow records}

The collecting of flow records is the starting point for constructing a network flow model. This step is crucial for obtaining accurate flow feature histograms and resulting mixture models. In this paper we assume that flow records are collected by the network equipment (hardware or software). The well known solutions allowing this are Cisco NetFlow, IPFIX and sFlow. The most common architecture of such concepts consist of two main components: an exporter that is responsible for creation of flow records from observed traffic and a collector that collects and processes flow records generated by the exporter. Collector is usually a software running on a commodity server.

Most importantly, packets must not be sampled by the exporter. Packet sampling techniques introduce bias to the collected data, which must be compensated for. The estimation of flows features histograms from sampled data was extensively studied in \cite{duffield2003estimating}, \cite{yang2009sample}, \cite{antunes2016estimation} and \cite{antunes2019regularized}, and falls outside scope of this paper. NetFlow and IPFIX exporters usually can operate in non-sampling mode. This is not always the case for sFlow, which sometimes imposes mandatory sampling. Therefore, before starting data collection, hardware or software must be configured to work in a non-sampling mode.

The software framework developed by us is designed to be used primarily on the top of the \texttt{nfdump} toolset\footnote{\url{http://github.com/phaag/nfdump}}. \texttt{nfdump} is an open-source framework used to collect and process flow records, created by Peter Haag. It supported preliminarily NetFlow data only, but extensions for processing other flow record formats were implemented afterwards. It stores flow records in nfcapd binary file format, which is the input format of our framework. This means that our framework can be used to analyze flow records collected in all formats supported by \texttt{nfdump}, which currently are: NetFlow v1, v5/v7, v9, IPFIX and sFlow (both IPv4 and IPv6).

Other important configuration parameters regarding the exporter are timeouts. In the case of NetFlow, these are:

\begin{itemize}

\item inactive timeout,

\item active timeout.

\end{itemize}

\emph{Inactive timeout} is the time, after which the particular flow record is exported under the condition that the exporting process does not collect any packet belonging to that flow. This means that packets with the same flow-defining key values, collected before reaching inactive timeout since the last packet will be considered as a single flow, and those collected after the timeout will be treated as a part of a new flow. The value of this timeout is a matter of flow definition and is crucial. For example, in the case of flowlet research, one would set this parameter to some subsecond value. This parameter strongly affects the resulting traffic model.

\emph{Active timeout} defines the time after which particular flow record is exported even when the flow is still active. The aim of this parameter is to limit the amount of memory required to store active flow records and counters in the exporter. In an ideal situation, it should not affect collected flow features as split records can be merged back (see below). However, the process of flows splitting performed by an exporter can introduce errors. Therefore, the timeout should be set to the highest value possible for particular hardware under expected load in order to reduce flow splitting as much as possible.

\newpage

\subsection{Cleaning the data}

Collected data must be groomed before performing the next steps. The exporter often generates flow records from multiple interfaces of a single device, so the same flow is reported as incoming flow on a one interface and as outgoing flow on another one. Moreover, it may happen that collector gathers data from multiple devices deployed within the same network, so the same flow may be reported multiple times by multiple exporters. Therefore, it must be ensured that data contains only flow records from a single interface of a single device. Moreover, we have noticed that NetFlow hardware exporters sometimes provide corrupted flow records characterized by implausible durations, which need to be filtered out as well.

The \texttt{nfdump} command line tool can be used to filter out flow records in the nfcapd format. It supports intuitive tcpdump-like filter expressions syntax and performs the operation pretty quickly. In case of single-homed network simple filtering of flow records that passes boarder router interface connected to the Internet can be applied. If a network is multi-homed, used filter should include all routers and interfaces that are connected upstream.



Wrong flow durations may not be only the results of a flow record corruption, but also an inevitable artifact caused by flow exporter. Several artifacts related to timing have been reported in literature \cite{hofstede2014flow}. Due to clock synchronization and precision issues, start and end times of flow records may not be precise or just simply incorrect. Another category of timing artifacts is the imprecise or erroneous expiration of flow records \cite{hofstede2013measurement}.


However, the calculation of flow duration requires accurate timestamps. Similarly, any time-related properties, like packet rate or bit rate, require precise timing. Therefore, used flow exporter must be examined whether it can provide accurate timestamps. In order to determine the reliability of flow timestamp records one can use the algorithm described in \cite{10.1007/978-3-662-43862-6_18}. In case of our data, we decided not to provide any time-related properties due to inaccuracy of used hardware NetFlow exporter. Ideally, packet-level traces should be analyzed instead flow-level records in order to obtain accurate flow rate and duration distributions.

\subsection{Merging of flow records}

Flows which were split into separate records due to the active timeout must be merged back into a single record in order to obtain accurate flow length, size or duration values. There is no software available to perform this operation so we developed a dedicated tool, available in our framework (called \texttt{merge}).

All flow records are processed in order of appearance. When a flow with a duration shorter than $active\_timeout - inactive\_timeout$ is encountered, it is dumped immediately as it is too short to be considered as a flow which was split due to active timeout. Initially, we were dumping flows with a duration lower than $active\_timeout$, but this was a pitfall because NetFlow agents do not always export active flow records accurately on active timeout. Instead, we found out that on our hardware the export of active flows starts after a delay approximately equal to $active\_timeout - inactive\_timeout$, so we used this value as a safe decision threshold. Research presented in \cite{hofstede2013measurement} shows that the \emph{active timeout} deviation can vary between different hardware or even firmware versions.

Flows with a duration greater than $active\_timeout - inactive\_timeout$ are considered as potential candidates to merge with subsequent flow records. Therefore, instead of being dumped, they are temporarily cached. When a new flow record with the same key is encountered, it is verified if the arrival time of the first packet is within the inactive timeout interval of the last packet of the cached flow. If so, these flow records are merged by summing their packets and octet counters and adjusting the first and last packet arrival times. Otherwise, the cached flow is dumped immediately and the new flow is either cached or dumped depending on the condition indicating if it can be considered as a candidate to merge. 

An additional lesson learned is that some portion of the flow records is erroneous. For example, a timestamp of the first packet of one flow is within the period when another flow with the same key is active (between its first and last packet timestamps). Such flow records are also filtered out by our merge tool. For example, in case of our campus data, erroneous flow records accounted for 0.0046\% of all collected records.

\subsection{Data binning}

Data binning is a step aimed at reducing the amount of data processed. We took advantage of the fact that next stages aimed at fitting and plotting do not have to be performed on complete flow records. Instead, they can operate on histograms (frequency distribution tables), calculated by binning flow records into buckets according to the selected parameter (such as flow length or size). Histogram files can also be easily published as they are many orders of magnitude smaller and, unlike flow records, do not contain private information such as IP addresses. We provide a tool called \texttt{hist} which performs flow binning and outputs histogram file in CSV format.

Binning the data into buckets of a width equal to one gives the most precise histogram. However, the resulting number of bins can be huge. This is especially problematic in the case of values of high granularity, such as flow sizes. Each distinct flow size results in a separate bucket entry, which means that for large flows, there are actually separate buckets for each flow. For example, in case of our campus data 4 billion flows resulted in 905 thousand separate buckets, yielding a 285 MB CSV file.

The solution to the above problem is to use bins of variable widths. For short and small flows, it is desirable to keep precise bins as they account for the vast majority of flows. However, precise bins are not essential for large flows. Therefore, logarithmic binning is the most appropriate scheme. Logarithmic binning can also significantly reduce size of histogram files. In our case it reduced the number of buckets to 44 thousand, which is a reduction by 95\%. Information loss introduced by it is negligible for the accuracy of fitted mixtures and CDF plots, however, it introduces distortion for PDF plots. Therefore, it must be appropriately compensated for during plotting.

\begin{figure}[!htb]
\centering
\includegraphics[scale=0.472]{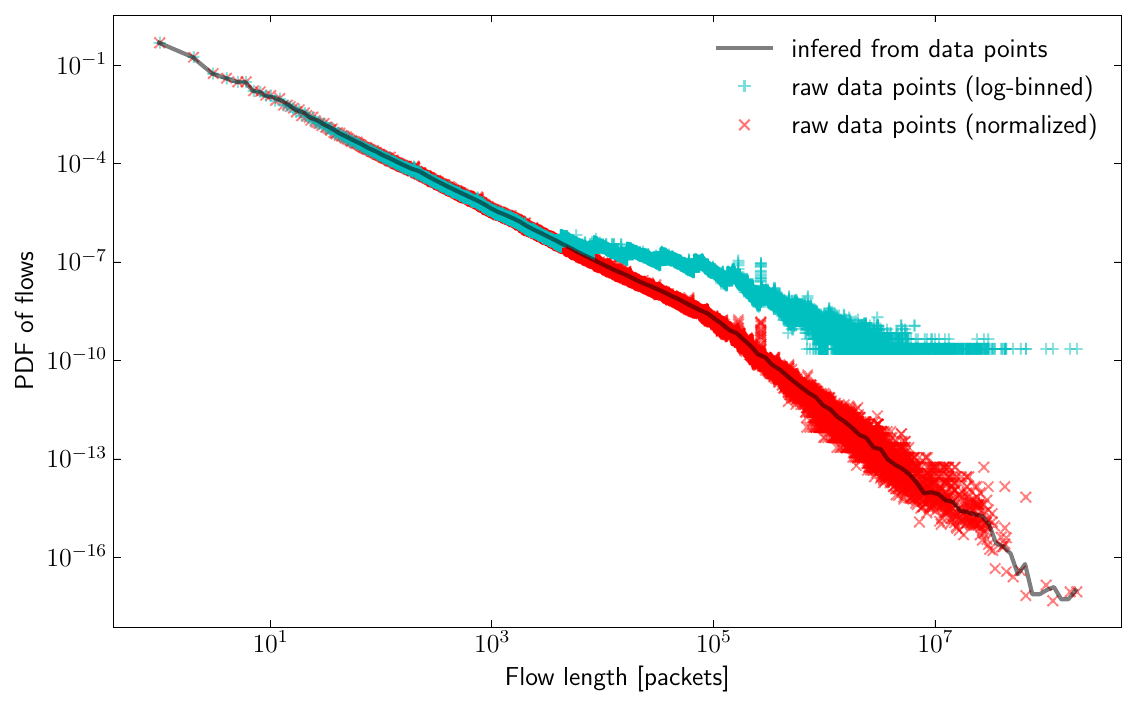}
\caption{Distortion of PDF points due to variable-width binning.}
\label{normalization-example}
\end{figure}

In the case of empirical PDF line plots, the best option is to first calculate the interpolated CDF, and second, to differentiate it in order to obtain the PDF line. Such an approach allows circumventing the distortion introduced by variable width binning. The problem appears with plots of PDF datapoints. The \autoref{normalization-example} presents the plot of PDF of flows in function of flow length. Data points are values of each bin of histogram. The solid line presents the PDF calculated by differentiating the CDF inferred from the data points. It can be seen that log-binned data points (cyan color) for bins wider than one are placed above their actual positions. This is because the variable-width binning bumps up number of flows in those bins. The correct way to compensate that boost is to divide sums in each bucket by the distance to the next non-empty bucket. This normalization procedure is implemented in the \texttt{plot} module in our framework, which can be used for plotting histograms, as well as fitted mixtures.

\subsection{Fitting of mixture models}

The fitting of the probability distribution to the series of the observed data is a process of finding a probability distribution and its parameters. However, due to the complexity of Internet traffic (many network applications and different users' behavior) single distribution cannot be fitted to match the collected data accurately. In such a situation, a mixture of distribution can be used. Such a model is a collection of other well known distributions called mixture components.

Finding mixture components and their weights is not a trivial process, especially when compared to the process of single distribution fitting where maximum-likelihood estimation (MLE) can be applied. To estimate the parameters of a statistical model composed of mixture components, a more sophisticated method must be used. One of the most commonly used machine learning algorithms for this purpose is the Expectation-Maximisation (EM) algorithm, which was also used by us. For details regarding the EM algorithm see \cite{EM} or our implementation \cite{github-flow-models}.

We have implemented the EM algorithm in a tool called \texttt{fit}, which is a part of our framework. It takes flow histogram CSV file as an input and performs distribution mixture fitting. In order to start the EM algorithm, an initial distribution mixture has to be provided. Its parameters are then iteratively refined in order to find the local optimum. Our tool can receive an initial distribution mixture from a user, but it can also generate an initial mixture for a particular dataset on its own, which means that a user has to only provide number and types of distributions used in mixture.

Currently, \emph{uniform}, \emph{normal}, \emph{lognormal}, \emph{Pareto}, \emph{Weibull} and \emph{gamma} distributions can be used in mixtures fitted by our tool. However, we have discovered, that \emph{uniform} and \emph{lognormal} distributions are usually sufficient to provide an accurate mixture model of flow lengths and sizes. They have an advantage of being fast to fit, since their maximization steps have analytical solutions, whereas some other distribution parameters (\emph{Weibull} or \emph{gamma}) must be calculated using numerical optimization methods. Another advantage is that they are widely implemented, so distribution mixtures composed of them can be usable in various network simulators and traffic generators.

Flow lengths and sizes are quantized values, so they should be approximated using discrete distributions. However, continuous distributions are considerably easier to model and use. In order to use continuous distributions for accurate modeling of a discrete data, it is required to properly handle them during fitting steps. Our implementation can be consulted for details on this \cite{github-flow-models}. Moreover, values generated using such models must be properly rounded, which is described in the next section.

It is more important to ensure accurate fitting for short and small flows, because they account for the majority of traffic. There are very few flows at the tail (which can be seen on PDF plots), so excessively accurate fitting to them would result in an overfitted model. For example, flows of length of one and two packets make up more than 50\% of total number of flows, since they are generated by special phenomena, like port scanning or DNS queries. Therefore, it is most beneficial to model such flows using \emph{uniform} distributions and use heavy-tailed distributions (like \emph{lognormal}) for longer flows. In case of our campus data, we discovered that modeling flows up to 6 packets with \emph{uniform} distributions and the rest with \emph{lognormal} gives the best results. Such a mixture is in fact a hybrid discrete-continuous mixture.


\subsection{Generating flows from models}

In order to be used for benchmarking of network mechanisms, models must enable the generation of traffic with distribution exactly matching that of the used models.

Firstly, \emph{flows} distribution mixture must be used to generate a random sample of flow length or size. A single distribution, randomly chosen from a mixture according to specified weights, has to be used to generate a random value. In case of the \texttt{scipy.stats} package, the \texttt{rvs} method can be used for this purpose. Next, in the case of non-continuous variables, such as flow length or size, the generated value must be rounded to the next nearest integer (by truncating the fractional part and adding one). In the case of flow size, values lower than the minimum packet size (64 bytes in the case of Ethernet) must be also rounded up to the minimum packet size.

As we have discovered, the average packet size depends on flow length/size. Thus, in order to accurately model the amount of traffic generated by flows of particular length, one has to obtain an average packet size that flow. To do this, the value of PDF of \emph{octets} at given sample point must be obtained (PDF should be calculated by differentiating the CDF of mixture). Value of PDF of \emph{octets} should be divided the value of PDF of \emph{packets}, obtained in a similar way. In the end, the resulting value has to be multiplied by the average packet size of the used model.

The tool \texttt{generate} in our framework can be used as a reference how to properly generate flows from distribution mixtures.

\section{Campus traffic model}

We applied the methodology described in the tutorial to the real traffic traces in order to extract models of flow lengths and sizes. We collected NetFlow records of all flows passing through the Internet-facing interface of the AGH University of Science and Technology wired network over 30 consecutive days. Flows traveling in both directions (upstream and downstream) were collected separately. We only collected dataplane traffic, without any control traffic (like OpenFlow messages between switches and controllers, etc.). Metadata of our dataset is presented in \autoref{tab-meta}. In total, we collected over 4 billion flows comprising 317 billion packets. The amount of transmitted data was over 275~TB. \autoref{tab-stats} presents statistics of the collected flows.

\begin{table}[!h]
\scriptsize
\caption{Metadata of collected flows}
\begin{center}
\begin{tabular}{@{}lrr@{}}
\toprule
\textbf{Dataset name} & agh\_2015 &  \\
\textbf{Flow definition} & 5-tuple &  \\
\textbf{Exporter} & Cisco router & (NetFlow) \\
\textbf{L2 technology} & Ethernet &  \\
\textbf{Sampling rate} & none &  \\
\textbf{Active timeout} & 300 & seconds \\
\textbf{Inactive timeout} & 15 & seconds \\
\textbf{Collection duration} & 30 & days \\
\bottomrule
\end{tabular}
\label{tab-meta}
\end{center}
\end{table}

\newpage

\begin{table}[!h]
\scriptsize
\caption{Statistics of collected flows}
\begin{center}
\begin{tabular}{@{}llrr@{}}
\toprule

\textbf{All traffic} & \textbf{Number of flows} & 4 032 376 751 & flows \\
& \textbf{Number of packets} & 316 857 594 090 & packets \\
& \textbf{Number of octets} & 275 858 498 994 998 & bytes \\
& \textbf{Average flow length} & 78.578370 & packets \\
& \textbf{Average flow size} & 68410.894128 & bytes \\
& \textbf{Average packet size} & 870.607188 & bytes \\

\midrule
\textbf{TCP-only} & \textbf{Number of flows} & 2 171 291 495 & flows \\
& \textbf{Number of packets} & 264 608 398 768 & packets \\
& \textbf{Number of octets} & 244 338 999 568 258 & bytes \\
& \textbf{Average flow length} & 121.866824 & packets \\
& \textbf{Average flow size} & 112531.643094 & bytes \\
& \textbf{Average packet size} & 923.398504 & bytes \\

\midrule
\textbf{UDP-only} & \textbf{Number of flows} & 1 737 523 167 & flows \\
& \textbf{Number of packets} & 50 730 510 162 & packets \\
& \textbf{Number of octets} & 31 098 128 445 645 & bytes \\
& \textbf{Average flow length} & 29.197027 & packets \\
& \textbf{Average flow size} & 17897.964779 & bytes \\
& \textbf{Average packet size} & 613.006421 & bytes \\

\bottomrule
\end{tabular}
\label{tab-stats}
\end{center}
\end{table}

Dormitories, populated with nearly 8000 students, generated 69\% of the traffic. The rest of the university (over 4000 employees) generated 31\%. In the case of dormitories, 91\% of traffic was downstream traffic (from the Internet).
In the case of rest of the university, downstream traffic made up 73\% of the total traffic. Therefore, statistics and models presented in this paper can also be considered as representative of residential traffic. The breakdown of observed traffic according to the origin is shown in \autoref{tab-source}.

\begin{table}[!h]
\scriptsize
\caption{Traffic shares by source (\%)}
\begin{center}
\begin{tabular}{@{}lrrrrrr@{}}
\toprule
& \multicolumn{3}{c}{\textbf{Dormitories}} & \multicolumn{3}{c}{\textbf{Rest of the campus}} \\
\cmidrule(lr){2-4} \cmidrule(l){5-7}
& all & down & up & all & down & up \\
\midrule
\multirow{2}{*}{\textbf{Flows}} & 41.62 & & & 58.38 & & \\
& & 50.05 & 49.95 & & 50.02 & 49.98 \\
\midrule
\multirow{2}{*}{\textbf{Packets}} & 67.72 & & & 32.28 & & \\
& & 63.88 & 36.12 & & 59.09 & 40.91 \\
\midrule
\multirow{2}{*}{\textbf{Octets}} & 68.66 & & & 31.34 & & \\
& & 91.41 & 8.59 & & 73.00 & 27.00 \\
\bottomrule
\end{tabular}
\label{tab-source}
\end{center}
\end{table}

The trace consists mostly of TCP and UDP traffic, which accounted for 54\% and 43\% of flows, respectively. However, TCP flows were responsible for nearly 89\% of transmitted data. Having in mind different nature of TCP and UDP traffic, in addition to the general model incorporating all observed flows (\textbf{All traffic}), we also provide models for \textbf{TCP-only} and \textbf{UDP-only} flows. Breakdown of traffic by the protocol is presented in \autoref{tab-protocol}.

\begin{table}[!h]
\scriptsize
\caption{Traffic shares by transport layer protocol (\%)}
\begin{center}
\begin{tabular}{@{}lrrr@{}}
\toprule
& \textbf{TCP} & \textbf{UDP} & \textbf{Other} \\
\midrule
\textbf{Flows} & 53.85 & 43.09 & 3.06 \\
\midrule
\textbf{Packets} & 83.51 & 16.01 & 0.48 \\
\midrule
\textbf{Octets} & 88.57 & 11.27 & 0.15 \\
\bottomrule
\end{tabular}
\label{tab-protocol}
\end{center}
\end{table}

In the subsequent paragraphs and figures we analyze the accuracy of the model developed and presented in this work. Namely, we compare the output of mixture model with data originally collected from the network and evaluate how well our model reflects real situation in the network.

Flows cumulative distribution function (CDF) tells what fraction of flows are flows of up to given length. Packets and octets CDFs tell what fraction of overall traffic is contributed by these flows. \autoref{tab-pct-length} and \autoref{tab-pct-size} presents selected values of empirical CDF derived from the collected data.

\begin{figure*}[!htb]
\centering
\includegraphics[scale=0.472]{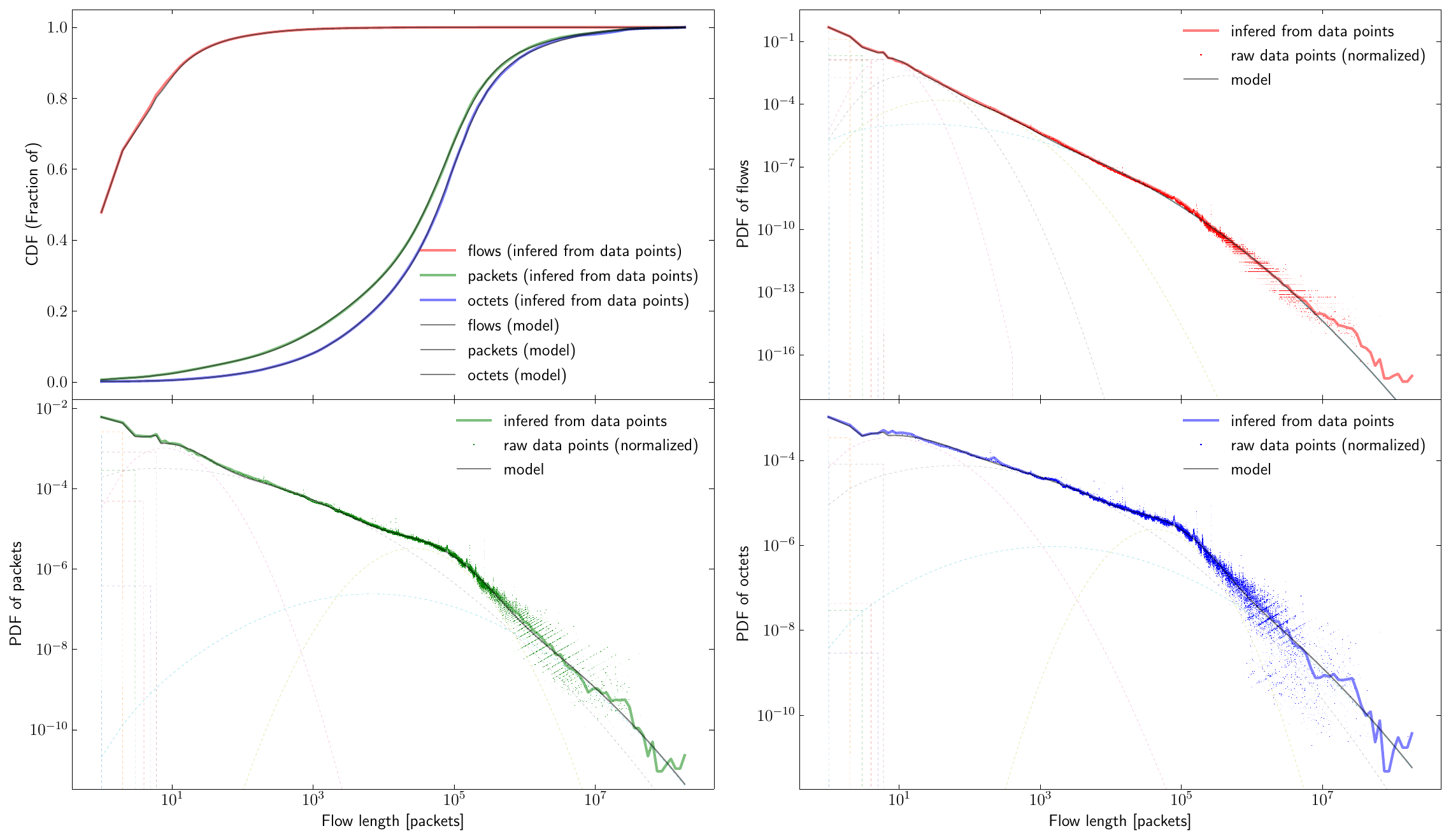}
\caption{Distribution plots in function of \textbf{flow length} (number of packets) (\textbf{All traffic}).}
\label{single-all-length}
\end{figure*}

\begin{figure*}[!htb]
\centering
\includegraphics[scale=0.472]{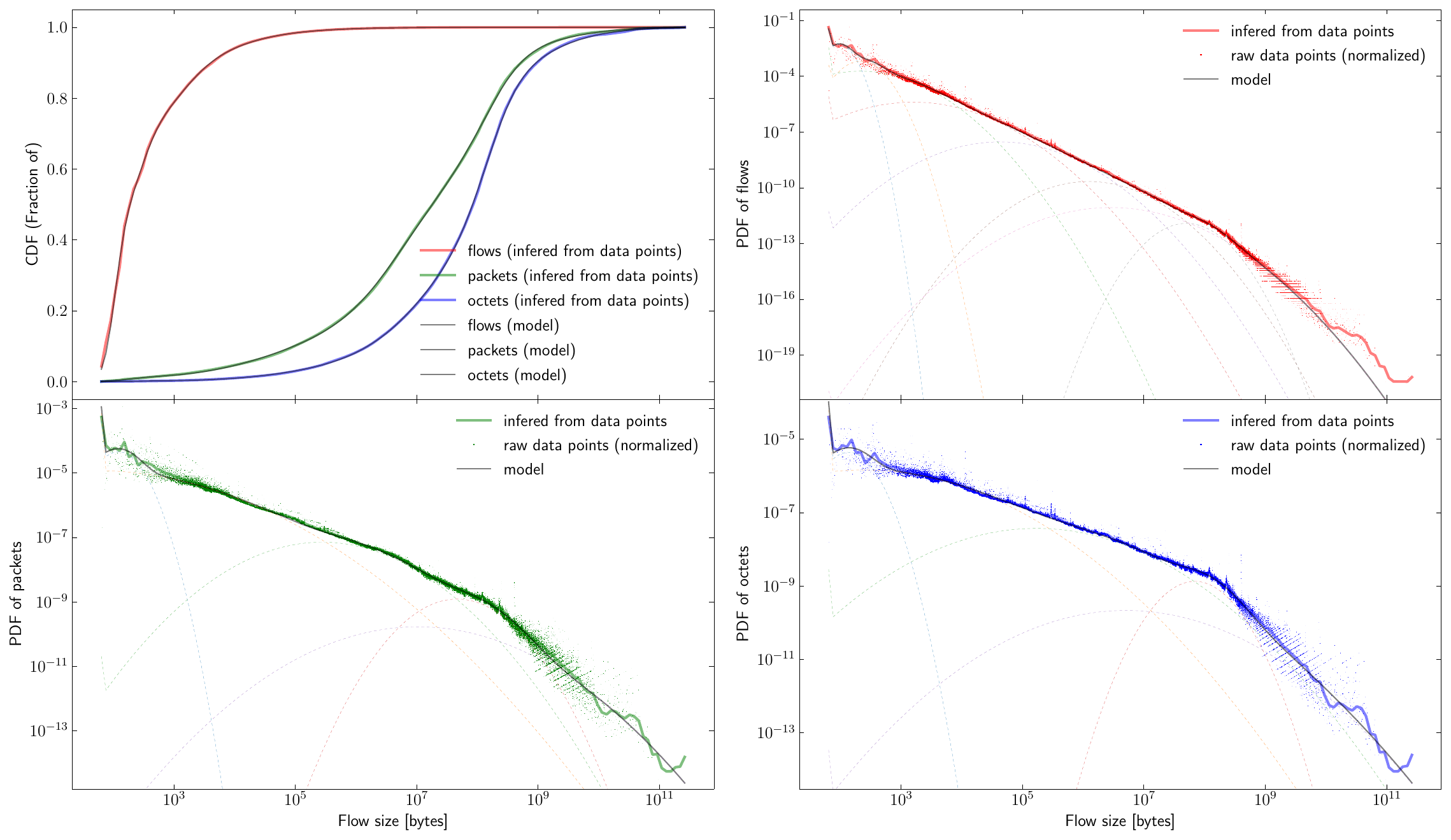}
\caption{Distribution plots in function of \textbf{flow size} (amount of bytes) (\textbf{All traffic}).}
\label{single-all-size}
\end{figure*}

\begin{figure*}[!htb]
\centering
\includegraphics[scale=0.472]{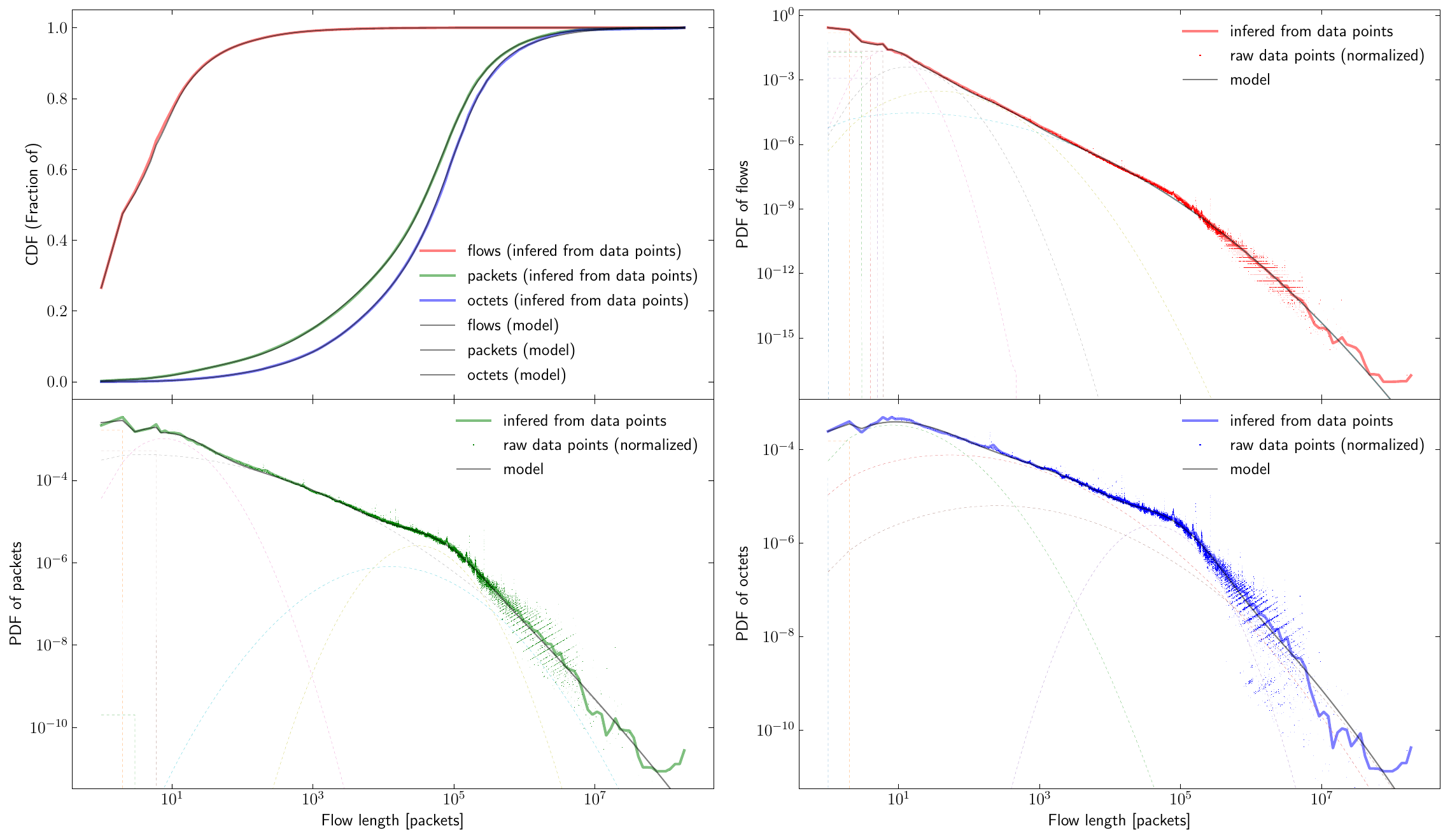}
\caption{Distribution plots in function of \textbf{flow length} (number of packets) (\textbf{TCP-only}).}
\label{single-tcp-length}
\end{figure*}

\begin{figure*}[!htb]
\centering
\includegraphics[scale=0.472]{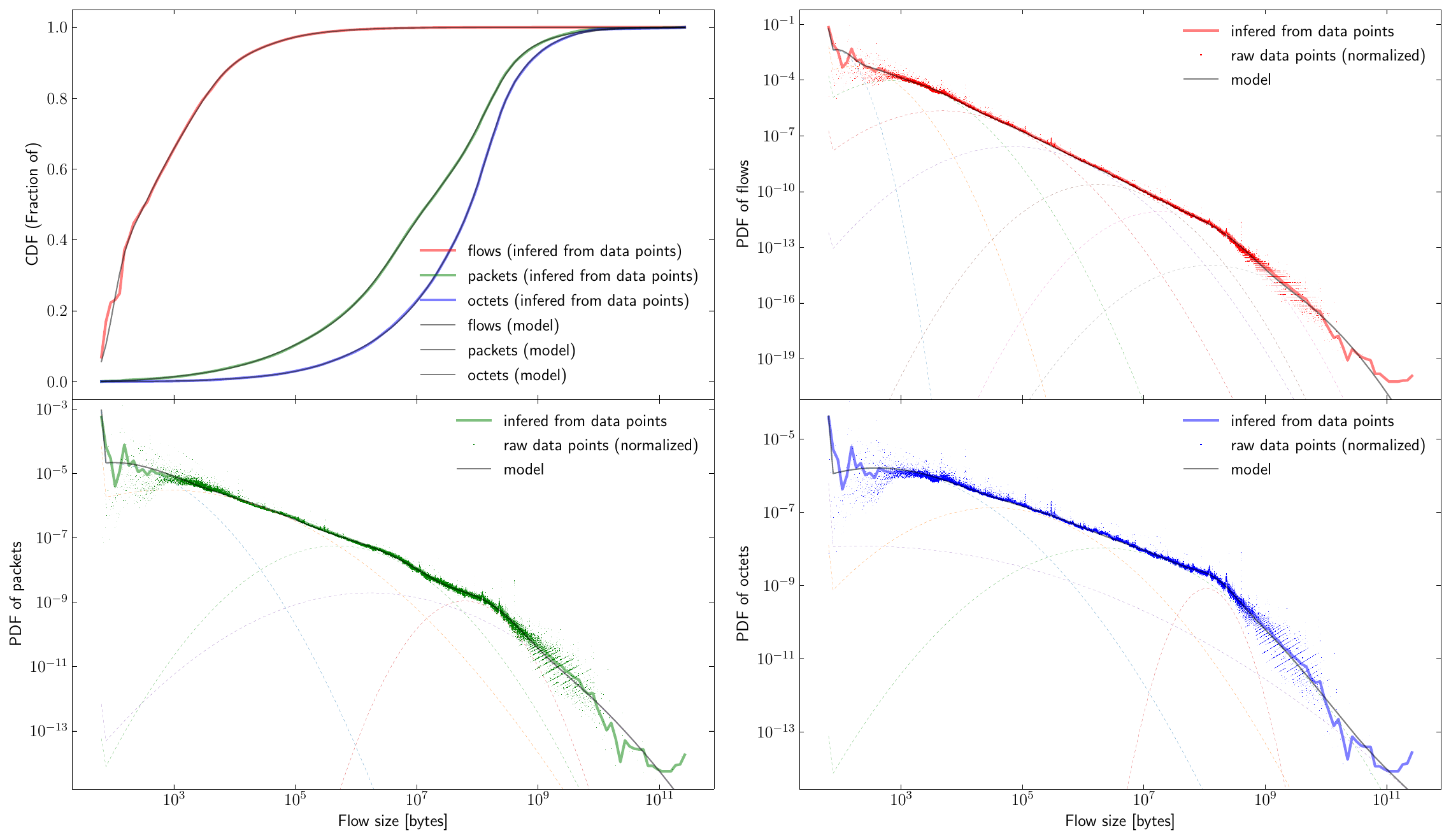}
\caption{Distribution plots in function of \textbf{flow size} (amount of bytes) (\textbf{TCP-only}).}
\label{single-tcp-size}
\end{figure*}

\begin{figure*}[!htb]
\centering
\includegraphics[scale=0.472]{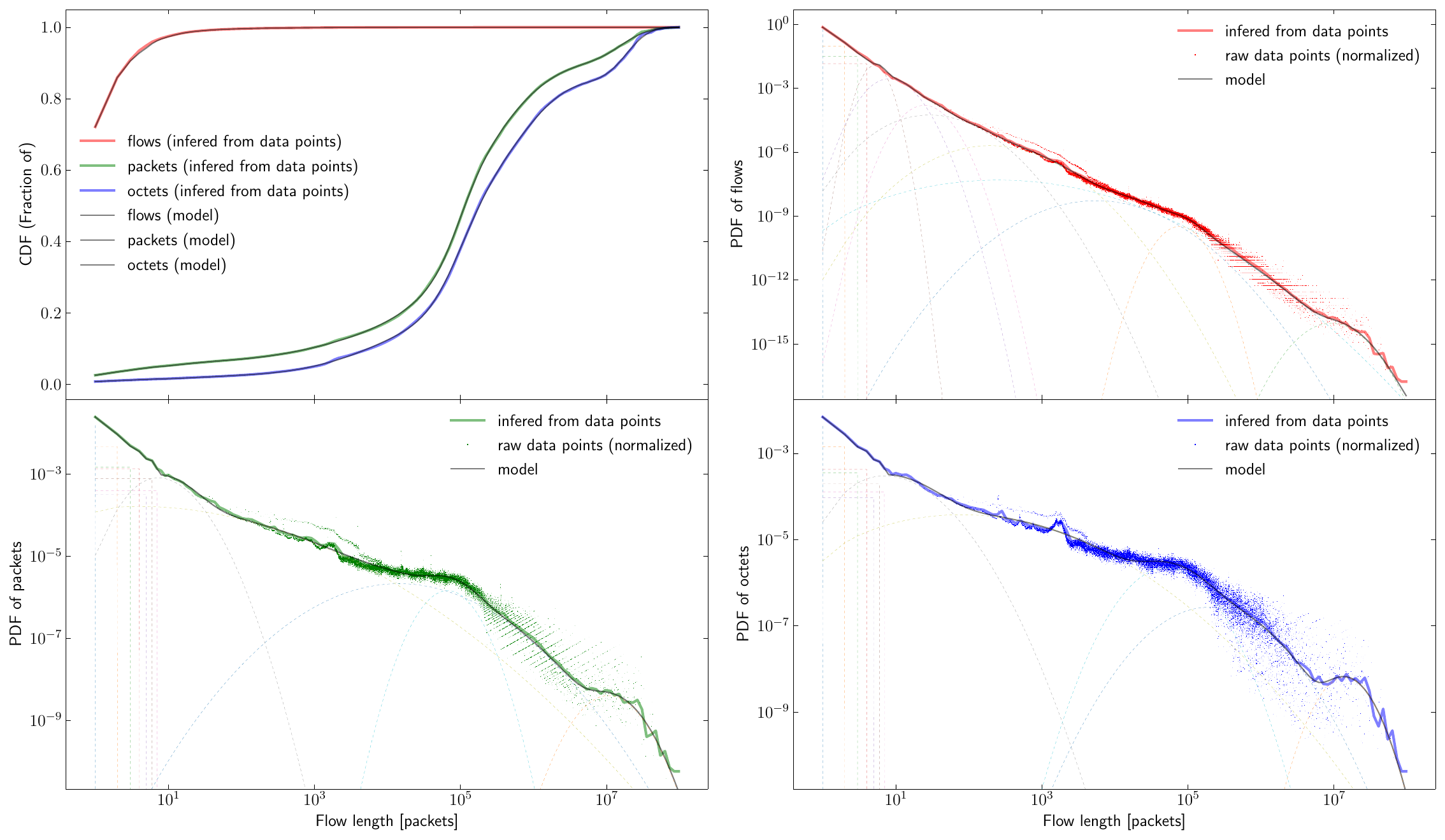}
\caption{Distribution plots in function of \textbf{flow length} (number of packets) (\textbf{UDP-only}).}
\label{single-udp-length}
\end{figure*}

\begin{figure*}[!htb]
\centering
\includegraphics[scale=0.472]{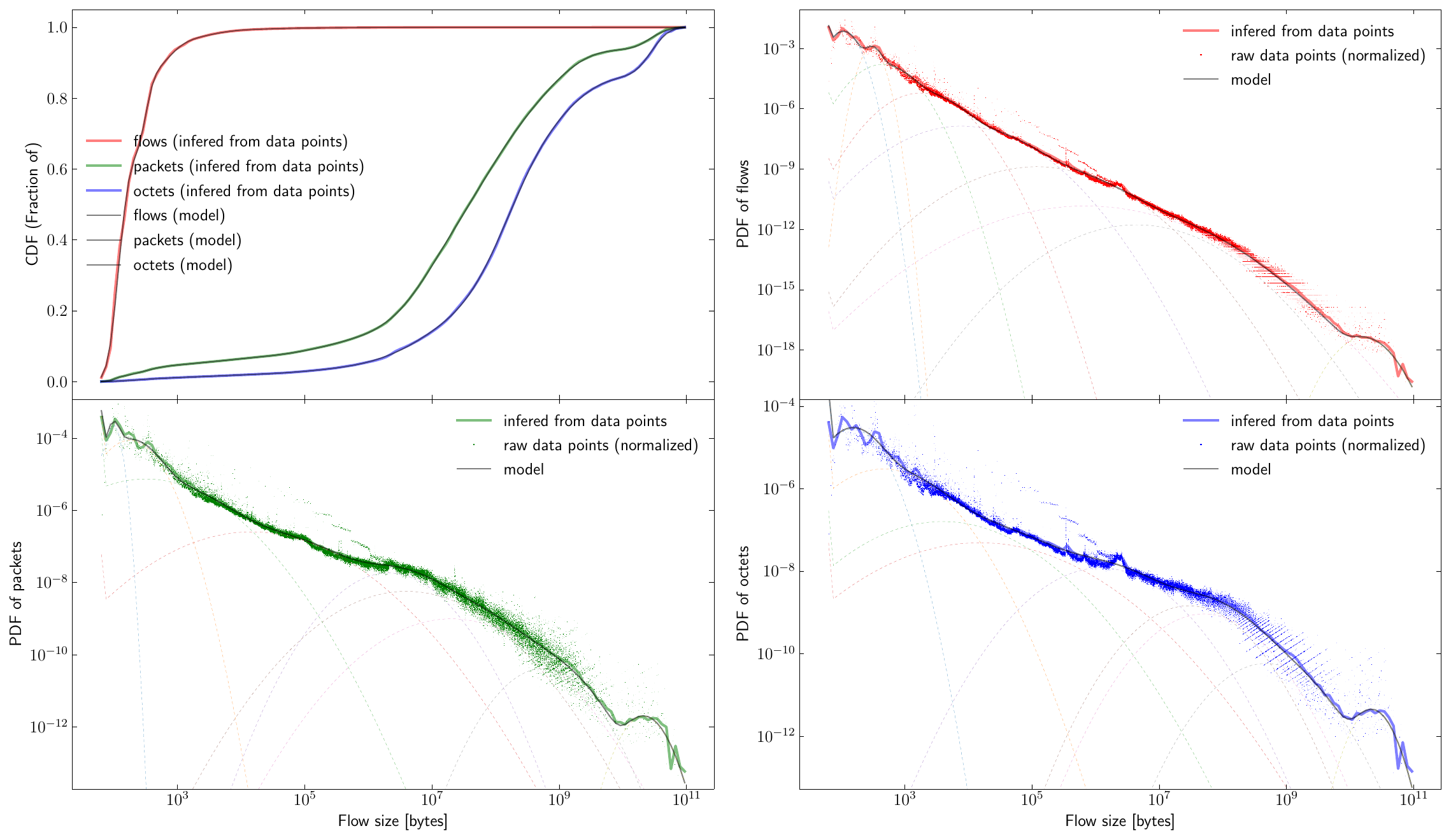}
\caption{Distribution plots in function of \textbf{flow size} (amount of bytes) (\textbf{UDP-only}).}
\label{single-udp-size}
\end{figure*}

\autoref{single-all-length}-\ref{single-udp-size} show flows, packets and octets distributions as functions of flow length and flow size, respectively. Starting from the upper left plot, CDFs are presented on a single plot. The next plots show probability density functions (PDF), each on a separate plot. The colored solid lines present CDFs and PDFs inferred from the data (empirical distribution functions). Rasterized datapoints are also visible on the plots. Black solid lines are CDFs and PDFs calculated from the fitted mixture models. Presenting them on the same figures as data allows graphical assessment of fitting accuracy. Moreover, on PDF plots, each mixture component is plotted as a dashed line.

\begin{figure*}[!htpb]
\centering
\includegraphics[scale=0.472]{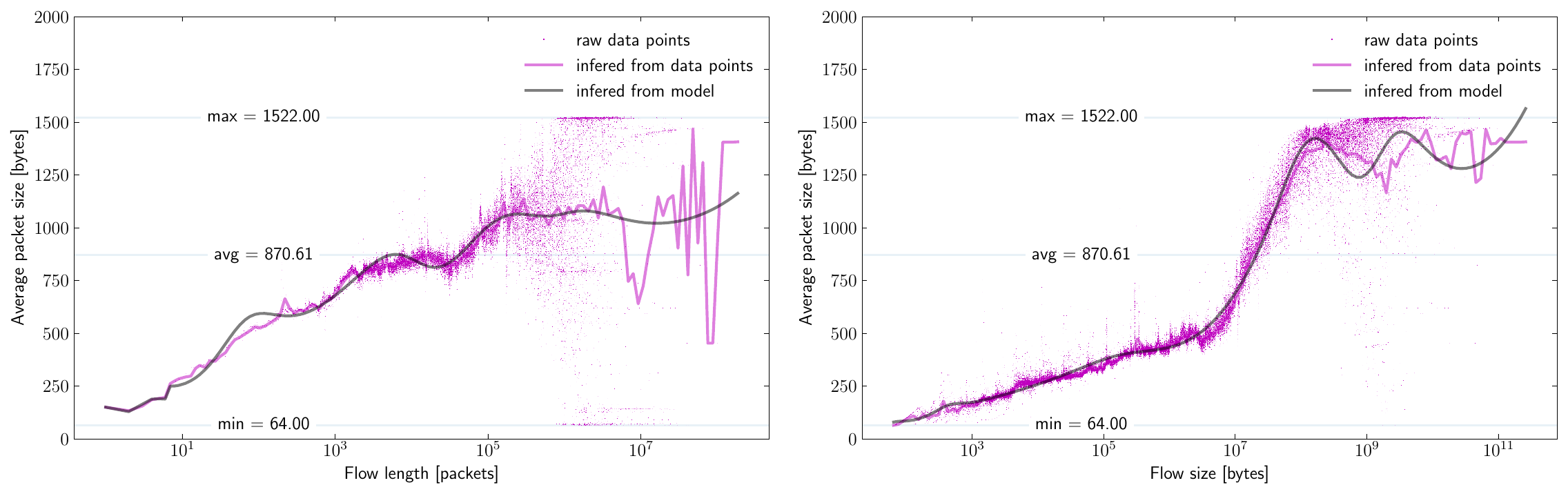}
\caption{Average packet size in functions of flow length and flow size (\textbf{All traffic}).}
\vspace{1cm}
\label{all-packet-sizes}
\end{figure*}

\begin{figure*}[!htpb]
\centering
\includegraphics[scale=0.472]{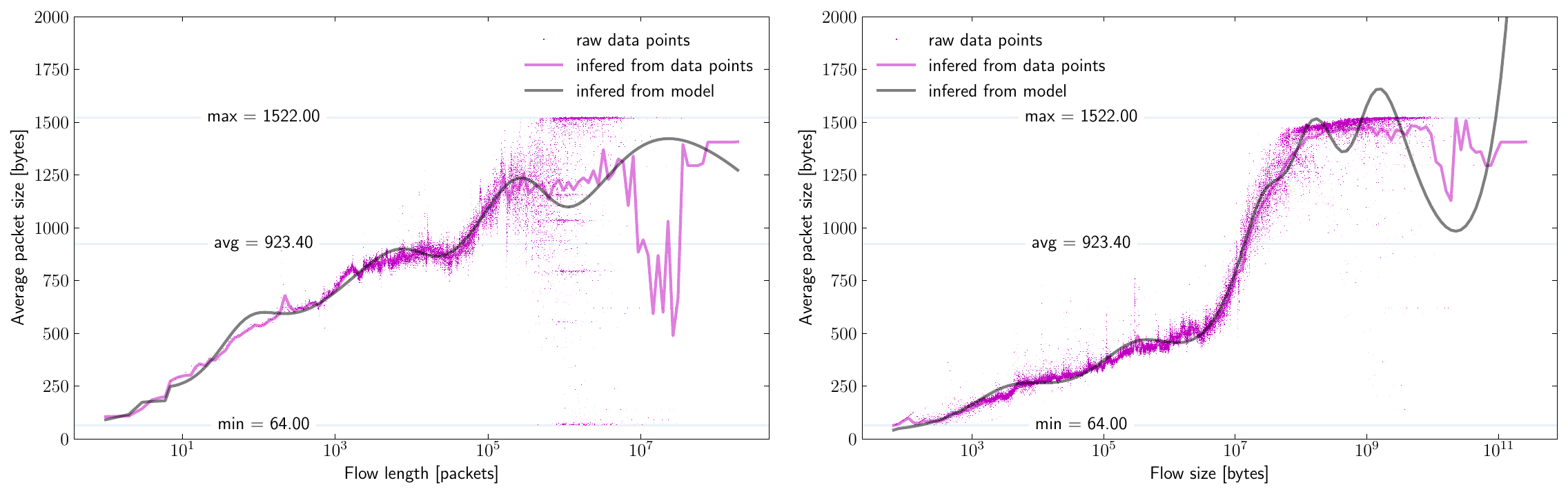}
\caption{Average packet size in functions of flow length and flow size (\textbf{TCP-only}).}
\vspace{1cm}
\label{tcp-packet-sizes}
\end{figure*}

\begin{figure*}[!htpb]
\centering
\includegraphics[scale=0.472]{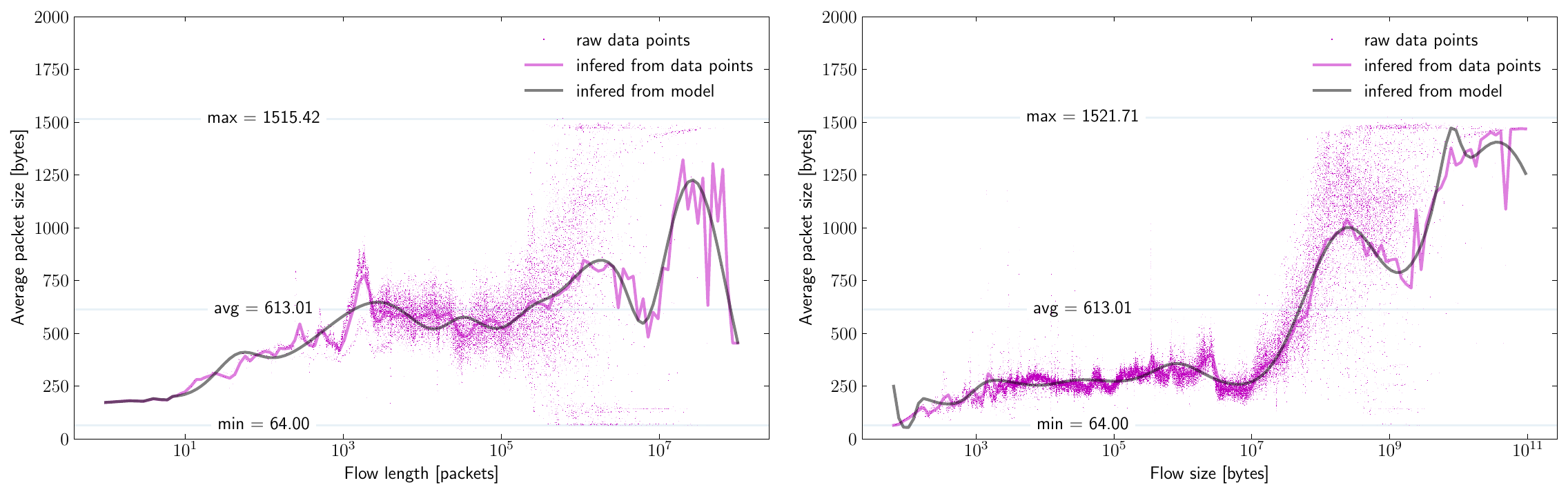}
\caption{Average packet size in functions of flow length and flow size (\textbf{UDP-only}).}
\label{udp-packet-sizes}
\end{figure*}

\autoref{all-packet-sizes}-\ref{udp-packet-sizes} presents average packet sizes for flows of particular lengths and sizes, respectively. It can be seen that longer/larger flows have a greater average packet size. This causes the horizontal gap between packet and octet CDFs, which can be seen in \autoref{single-all-length}-\ref{single-udp-size}. If the average packet size was constant and independent from flow length/size, packet and octet CDFs (green and blue lines) would overlap. The minimum packet size was 64 bytes and the maximum packet size was 1522 bytes, which means that IEEE 802.1Q VLAN tagged Ethernet frames were observed in the trace.

\begin{table*}[!h]
\scriptsize
\caption{Selected values of empirical distributions in function of \textbf{flow length}}
\begin{center}
\begin{tabular}{@{}lrrrrrrrrr@{}}
\toprule
\multirowcell{4}{\textbf{Flows of length up to} \\ (packets)} & \multicolumn{3}{c}{\textbf{All traffic}} & \multicolumn{3}{c}{\textbf{TCP-only}} & \multicolumn{3}{c}{\textbf{UDP-only}} \\
\cmidrule(l){2-10}
& \multicolumn{3}{c}{\textbf{Make up \%}} & \multicolumn{3}{c}{\textbf{Make up \%}} & \multicolumn{3}{c}{\textbf{Make up \%}} \\
\cmidrule(lr){2-4} \cmidrule(lr){5-7} \cmidrule(l){8-10}
& of flows & of packets & of octets & of flows & of packets & of octets & of flows & of packets & of octets \\
\midrule
1 & 47.8326 & 0.6087 & 0.1047            & 26.5673 & 0.2180 & 0.0248  &      72.0893 & 2.4691 & 0.6936 \\
2 & 65.3421 & 1.0544 & 0.1728            & 47.5666 & 0.5626 & 0.0652  &      85.9202 & 3.4165 & 0.9729 \\
4 & 74.8933 & 1.4696 & 0.2537            & 58.9241 & 0.8847 & 0.1226  &      93.4510 & 4.2818 & 1.2308 \\
8 & 84.1319 & 2.1958 & 0.4412            & 73.1538 & 1.6098 & 0.3050  &      96.9092 & 4.9940 & 1.4527 \\
16 & 90.5756 & 3.1633 & 0.7830           & 83.8696 & 2.6480 & 0.6583  &      98.4129 & 5.5985 & 1.7030 \\
32 & 94.3448 & 4.2601 & 1.2432           & 90.2560 & 3.8485 & 1.1430  &      99.1112 & 6.1370 & 1.9661 \\
64 & 96.4556 & 5.4778 & 1.8863           & 93.8820 & 5.1985 & 1.8303  &      99.4607 & 6.6728 & 2.2622 \\
128 & 97.7421 & 6.9597 & 2.7813          & 96.1150 & 6.8565 & 2.7917  &      99.6457 & 7.2502 & 2.6369 \\
256 & 98.5940 & 8.9276 & 4.1445          & 97.5940 & 9.0582 & 4.2560  &      99.7673 & 8.0068 & 3.2041 \\
512 & 99.1268 & 11.3677 & 5.8630         & 98.5178 & 11.7861 & 6.1010  &     99.8428 & 8.9430 & 3.9217 \\
1024 & 99.4631 & 14.4387 & 8.1638        & 99.0930 & 15.1682 & 8.5593  &     99.8988 & 10.3430 & 4.9581 \\
2048 & 99.6735 & 18.2982 & 11.5607       & 99.4477 & 19.3539 & 12.0957  &    99.9375 & 12.2954 & 7.1653 \\
4096 & 99.7996 & 22.9134 & 15.7892       & 99.6647 & 24.4758 & 16.6297  &    99.9562 & 14.1350 & 8.9254 \\
8192 & 99.8763 & 28.5198 & 21.1283       & 99.7966 & 30.6908 & 22.3573  &    99.9681 & 16.5167 & 11.1823 \\
16384 & 99.9249 & 35.6815 & 28.1727      & 99.8792 & 38.5230 & 29.8534  &    99.9772 & 20.1770 & 14.6856 \\
32768 & 99.9581 & 45.4696 & 37.5909      & 99.9345 & 49.0188 & 39.7829  &    99.9849 & 26.3610 & 20.1202 \\
65536 & 99.9805 & 58.5711 & 50.6240      & 99.9706 & 62.6382 & 53.3185  &    99.9915 & 36.8420 & 29.2514 \\
131072 & 99.9933 & 73.2594 & 67.2425     & 99.9902 & 77.0380 & 70.2250  &    99.9967 & 53.2005 & 43.7028 \\
262144 & 99.9981 & 83.9254 & 80.4537     & 99.9973 & 87.2900 & 83.5018  &    99.9988 & 66.2152 & 56.4890 \\
524288 & 99.9994 & 89.9609 & 87.8104     & 99.9992 & 92.7675 & 90.5868  &    99.9996 & 75.2048 & 65.9873 \\
1048576 & 99.9998 & 93.8090 & 92.4484    & 99.9998 & 96.0066 & 94.7181  &    99.9999 & 82.2277 & 74.5952 \\
2097152 & 99.9999 & 96.1224 & 95.3628    & 99.9999 & 97.9366 & 97.2627  &    99.9999 & 86.5530 & 80.3917 \\
4194304 & 100.0000 & 97.4921 & 97.1076   & 100.0000 & 99.0591 & 98.8117 &    100.0000 & 89.2491 & 83.6806 \\
8388608 & 100.0000 & 98.2356 & 97.9084   & 100.0000 & 99.4960 & 99.4221 &    100.0000 & 91.6152 & 86.0016 \\
16777216 & 100.0000 & 98.9065 & 98.5951  & 100.0000 & 99.6210 & 99.5394 &    100.0000 & 95.1656 & 91.1591 \\

\bottomrule
\end{tabular}
\label{tab-pct-length}
\end{center}
\end{table*}

\begin{table*}[!h]
\scriptsize
\caption{Selected values of empirical distributions in function of \textbf{flow size}}
\begin{center}
\begin{tabular}{@{}lrrrrrrrrr@{}}
\toprule
\multirowcell{4}{\textbf{Flows of size up to} \\ (bytes)} & \multicolumn{3}{c}{\textbf{All traffic}} & \multicolumn{3}{c}{\textbf{TCP-only}} & \multicolumn{3}{c}{\textbf{UDP-only}} \\
\cmidrule(l){2-10}
& \multicolumn{3}{c}{\textbf{Make up \%}} & \multicolumn{3}{c}{\textbf{Make up \%}} & \multicolumn{3}{c}{\textbf{Make up \%}} \\
\cmidrule(lr){2-4} \cmidrule(lr){5-7} \cmidrule(l){8-10}
& of flows & of packets & of octets & of flows & of packets & of octets & of flows & of packets & of octets \\
\midrule
64 & 4.3082 & 0.0548 & 0.0040   &             6.9223 & 0.0568 & 0.0039     & 1.1258 & 0.0386 & 0.0040 \\
128 & 32.3376 & 0.4196 & 0.0424   &           24.8657 & 0.2132 & 0.0168    & 38.9641 & 1.3367 & 0.2140 \\
256 & 56.8711 & 0.9477 & 0.1030   &           46.6519 & 0.5932 & 0.0486    & 67.4629 & 2.5868 & 0.4897 \\
512 & 71.1101 & 1.4143 & 0.1780   &           56.3675 & 0.8823 & 0.0812    & 87.9533 & 3.9487 & 0.8905 \\
1024 & 79.0397 & 1.9054 & 0.2622   &          65.9424 & 1.3192 & 0.1443    & 94.1627 & 4.6973 & 1.1340 \\
2048 & 85.1875 & 2.5299 & 0.3934   &          75.0551 & 1.9595 & 0.2634    & 96.9302 & 5.2317 & 1.3528 \\
4096 & 89.7285 & 3.3288 & 0.5845   &          82.3581 & 2.8099 & 0.4505    & 98.2762 & 5.7397 & 1.5681 \\
8192 & 93.3548 & 4.3316 & 0.8890   &          88.4784 & 3.9059 & 0.7634    & 99.0219 & 6.2580 & 1.8046 \\
16384 & 95.5706 & 5.4696 & 1.2613   &         92.2810 & 5.1607 & 1.1520    & 99.4008 & 6.7929 & 2.0457 \\
32768 & 96.9521 & 6.8385 & 1.7288   &         94.6877 & 6.6825 & 1.6473    & 99.5938 & 7.3776 & 2.2938 \\
65536 & 97.9398 & 8.6723 & 2.4034   &         96.4164 & 8.7216 & 2.3649    & 99.7202 & 8.1431 & 2.6245 \\
131072 & 98.6472 & 11.0558 & 3.3586   &       97.6511 & 11.3448 & 3.3790   & 99.8117 & 9.2470 & 3.0959 \\
262144 & 99.1008 & 13.7882 & 4.5873   &       98.4430 & 14.3563 & 4.6832   & 99.8686 & 10.4009 & 3.6807 \\
524288 & 99.4222 & 17.2034 & 6.3000   &       99.0041 & 18.1008 & 6.4993   & 99.9084 & 11.9168 & 4.4992 \\
1048576 & 99.6168 & 21.4048 & 8.4002   &      99.3427 & 22.6740 & 8.7190   & 99.9345 & 14.0095 & 5.5777 \\
2097152 & 99.7534 & 26.8008 & 11.3725   &     99.5790 & 28.4750 & 11.8410  & 99.9548 & 17.1643 & 7.3150 \\
4194304 & 99.8481 & 34.0396 & 15.4253   &     99.7403 & 36.0126 & 16.0511  & 99.9726 & 22.7803 & 10.1026 \\
8388608 & 99.9061 & 41.9744 & 20.4050   &     99.8405 & 43.9471 & 21.2798  & 99.9817 & 30.6529 & 13.1125 \\
16777216 & 99.9417 & 49.1419 & 26.5333   &    99.9015 & 50.9127 & 27.6685  & 99.9879 & 38.9022 & 17.2073 \\
33554432 & 99.9659 & 56.6984 & 34.8628   &    99.9425 & 58.1078 & 36.2686  & 99.9926 & 48.4453 & 23.4513 \\
67108864 & 99.9811 & 64.7532 & 45.4070   &    99.9683 & 66.0042 & 47.1078  & 99.9958 & 57.4070 & 31.7235 \\
134217728 & 99.9918 & 74.5103 & 60.2620   &   99.9864 & 76.1026 & 62.3844  & 99.9979 & 65.6134 & 43.3810 \\
268435456 & 99.9974 & 84.0454 & 75.4282   &   99.9958 & 85.9574 & 77.8295  & 99.9992 & 73.7219 & 56.4785 \\
536870912 & 99.9992 & 90.1133 & 84.9561   &   99.9988 & 91.9435 & 87.2988  & 99.9997 & 80.3462 & 66.4994 \\
1073741824 & 99.9998 & 93.8565 & 90.6774  &   99.9996 & 95.3848 & 92.7713  & 99.9999 & 85.7396 & 74.1849 \\

\bottomrule
\end{tabular}
\label{tab-pct-size}
\end{center}
\end{table*}

Finally, we provide mixture models fitting the distributions. They consist of \emph{uniform} and \emph{lognormal} distributions. The mixtures along with their parameters are provided in the \shortref{appendix} and in our GitHub repository \cite{github-flow-models}.

The numbers presented in tables are in line with the limited histograms of CAIDA and BME traces presented in \cite{megyesi2013analysis}. The CAIDA traces contain traffic recorded on a backbone link between Chicago and San Jose, whereas the BME trace represents traffic of Budapest University of Technology and Economics WAN interface. This confirms that our models indeed represent a realistic Internet traffic.

The provided plots and models can be used to evaluate various flow-oriented networking solutions. For example, we can use them to evaluate a traffic engineering mechanism based on heuristics that differentiates traffic into elephant and mice, similar to the one described in the \fullref{introduction}.

It can be seen that flows up to 1000 packets comprise 99.45\% of all flows in the network, but they are responsible only for roughly 8\% of transmitted data. Such an observation can be utilized to reduce management overheads and flow table size requirements in SDN. A traffic engineering system interested only in elephant flows larger than 1000 packets would have to process only 0.55\% of layer-4 flows present in the network, but would still cover 92\% of the overall amount of traffic. As knowledge about TCP/UDP headers is available in any monitoring tool, significant advantages may be achieved without using any deep packet inspection tools.

In case of more advanced solutions, distribution mixtures can be used to calculate their performance analytically. Ultimately, provided models can be used to generate traffic in network simulators and emulators to evaluate performance of flow-based TE solutions on network-scale.

\section{Conclusion}

The contribution of this paper is fourfold. Firstly, it provides a complete tutorial on methodology aimed at constructing network flow models from flow records.

Secondly, a ready-to-use and scalable framework implementing this methodology is published as an open source software. Due to applying big data techniques it scales horizontally and can be used to process an unlimited number of flow records and fit distribution mixtures to them.

Thirdly, the paper presents an example of applying the methodology to analyze flow records collected at the Internet-facing interface of the campus network. Flows number, and total sum of packet and octet distributions are extracted, analyzed and modeled as functions of both flow length and flow size. Models are represented by complete distribution mixtures provided with parameters. They are based on billions of flows which is considerably more comparing to the previous analyses mostly comprising tens of millions of flows and can be treated as an approximation of the general Internet traffic.

Last but not least, the presented models can be utilized as a unified benchmark enabling the comparable assessment of novel flow-oriented solutions, algorithms, and concepts. This especially applies to next-generation SDN techniques based on flow forwarding, like OpenFlow or P4. It makes our work timely and relevant as these next-generation flow-based approaches have been gaining attention over recent years. There is no other similar work giving such a reusable contribution. The expected impact is probably unlimited because each novel network flow-oriented mechanism should be validated under realistic assumptions regarding the system load.

\pagebreak

\section*{Acknowledgment}
\noindent The research was carried out with the support of the project "Intelligent management of traffic in multi-layer Software-Defined Networks" founded by the Polish National Science Centre under project no. 2017/25/B/ST6/02186.

\bigskip

\noindent The authors would like to thank Bogusław Juza for providing NetFlow flow record dumps.

\bibliography{./bib}
\bibliographystyle{elsarticle-num}

\onecolumn
\appendix
\section{Mixture models}
\label{appendix}

\noindent Each mixture is provided as a JSON object. The \emph{sum} field is the sum of the flows/packets/octets in the dataset. These sum values can be used for the calculation of the average packet size from the provided distribution mixture models. The \emph{mix} field contains a list of distribution components which form a mixture. Each component is represented by a list, containing its weight, the \texttt{scipy.stats} distribution name and distribution parameters. For generating samples from the provided distribution mixtures, one should follow the methodology described in the tutorial or use the \texttt{generate} tool provided in our framework.

\subsection{All traffic}
\subsubsection{Flows (length)}
\begin{lstlisting}[language=json]
{
  "sum": 4032376751,
  "mix": [
    [0.3050265769901237, "uniform", [0, 1]],
    [0.2484198800441619, "uniform", [0, 2]],
    [0.06366063664158106, "uniform", [0, 3]],
    [0.049216499659328734, "uniform", [0, 4]],
    [0.00931559166293732, "uniform", [0, 5]],
    [0.08217474157187248, "uniform", [0, 6]],
    [0.13126374982514846, "lognorm", [0.5207023493412835, 0, 7.8055992790704085]],
    [0.07328615421743477, "lognorm", [0.7701056575379265, 0, 22.10972501544739]],
    [0.0292891264871594, "lognorm", [1.1252645297342514, 0, 128.6451515069839]],
    [0.00834704290025077, "lognorm", [1.98383694524085, 0, 1084.4707584768782]]
  ]
}
\end{lstlisting}

\subsubsection{Packets (length)}
\begin{lstlisting}[language=json]
{
  "sum": 316857594090,
  "mix": [
    [0.002110368287712165, "uniform", [0, 1]],
    [0.005204902009087457, "uniform", [0, 2]],
    [0.0008611208380640299, "uniform", [0, 3]],
    [0.0001953848185699079, "uniform", [0, 4]],
    [1.9081822832806864e-06, "uniform", [0, 5]],
    [0.0048883553554204375, "uniform", [0, 6]],
    [0.02710154435301931, "lognorm", [0.9402388599292199, 0, 17.389638035271467]],
    [0.46884288920105865, "lognorm", [2.683257909795451, 0, 7959.875774748725]],
    [0.3938789200857053, "lognorm", [1.0517807783252187, 0, 78648.30394527224]],
    [0.09691460686907521, "lognorm", [2.1747855564359075, 0, 788388.3862527548]]
  ]
}
\end{lstlisting}

\subsubsection{Octets (length)}
\begin{lstlisting}[language=json]
{
  "sum": 275858498994998,
  "mix": [
    [0.0005522188937705094, "uniform", [0, 1]],
    [0.0006733419371011684, "uniform", [0, 2]],
    [8.712450629315575e-08, "uniform", [0, 3]],
    [1.7242898989451478e-07, "uniform", [0, 4]],
    [1.4615786843822656e-08, "uniform", [0, 5]],
    [0.0004861299541601516, "uniform", [0, 6]],
    [0.025410773535068393, "lognorm", [1.4734220306634367, 0, 60.469099992767916]],
    [0.47888389261362085, "lognorm", [2.3777777137900578, 0, 17998.560650283638]],
    [0.3478874901495348, "lognorm", [0.918261689867093, 0, 97153.89719008311]],
    [0.14610587874746267, "lognorm", [2.396148205892031, 0, 458689.5696009558]]
  ]
}
\end{lstlisting}

\pagebreak

\subsubsection{Flows (size)}
\begin{lstlisting}[language=json]
{
  "sum": 4032376751,
  "mix": [
    [0.44966670086340027, "lognorm", [0.3417219351724571, 0, 106.166372352337]],
    [0.20220241315279464, "lognorm", [0.5104895027884305, 0, 277.63774474539736]],
    [0.2686509806417942, "lognorm", [1.2406356394188367, 0, 1000.5544200478133]],
    [0.06647780135323868, "lognorm", [1.4509439390336445, 0, 13006.098381673599]],
    [0.010998574523761361, "lognorm", [1.3716500213320844, 0, 279076.8101102302]],
    [0.001570957282952998, "lognorm", [1.1704125831664687, 0, 4886421.753135197]],
    [0.000292175353983837, "lognorm", [1.5101192537041566, 0, 28081955.675893098]],
    [0.00014039682805832134, "lognorm", [0.7082820543792429, 0, 83124658.11210157]]
  ]
}
\end{lstlisting}

\subsubsection{Packets (size)}
\begin{lstlisting}[language=json]
{
  "sum": 316857594090,
  "mix": [
    [0.009407248052942033, "lognorm", [0.5695232010956377, 0, 168.31994829953425]],
    [0.17317893304105525, "lognorm", [2.7598572109187933, 0, 93971.53306040031]],
    [0.4397773291615796, "lognorm", [1.7856596583166613, 0, 6774586.520568432]],
    [0.2838759525683184, "lognorm", [1.096084781703366, 0, 154523629.11502388]],
    [0.09376053717610511, "lognorm", [2.143017425155127, 0, 1029106790.6910875]]
  ]
}
\end{lstlisting}

\subsubsection{Octets (size)}
\begin{lstlisting}[language=json]
{
  "sum": 275858498994998,
  "mix": [
    [0.0011947871676142936, "lognorm", [0.6394221616703024, 0, 205.47201866172307]],
    [0.06625789808407466, "lognorm", [2.595354698521618, 0, 202947.16944464625]],
    [0.5401266530263714, "lognorm", [2.3185516152408274, 0, 37006202.73334796]],
    [0.298097741735126, "lognorm", [0.8770896315727511, 0, 149441408.23497528]],
    [0.09432291998681219, "lognorm", [2.325205564350233, 0, 1137154460.9830294]]
  ]
}
\end{lstlisting}

\pagebreak

\subsection{TCP-only}
\subsubsection{Flows (length)}
\begin{lstlisting}[language=json]
{
  "sum": 2171291495,
  "mix": [
    [0.058043754929076714, "uniform", [0, 1]],
    [0.3105249919828018, "uniform", [0, 2]],
    [0.05509731658732791, "uniform", [0, 3]],
    [0.04775404343193455, "uniform", [0, 4]],
    [0.005882528205165834, "uniform", [0, 5]],
    [0.12529112884696628, "uniform", [0, 6]],
    [0.20654328341287823, "lognorm", [0.5122381561489622, 0, 7.9543319950074585]],
    [0.1218958369564864, "lognorm", [0.7507627341686453, 0, 21.74157327774004]],
    [0.052507436335709035, "lognorm", [1.1085442272126311, 0, 115.96889383386637]],
    [0.016459679311651095, "lognorm", [1.9932118957877463, 0, 845.9121788017795]]
  ]
}
\end{lstlisting}

\subsubsection{Packets (length)}
\begin{lstlisting}[language=json]
{
  "sum": 264608398768,
  "mix": [
    [2.2755551196918487e-34, "uniform", [0, 1]],
    [0.003314106999833896, "uniform", [0, 2]],
    [6.013523396795704e-10, "uniform", [0, 3]],
    [2.932420026100149e-12, "uniform", [0, 4]],
    [1.067649619016758e-11, "uniform", [0, 5]],
    [0.003146634905134179, "uniform", [0, 6]],
    [0.027932504515234154, "lognorm", [0.9545700860476148, 0, 17.675326565099915]],
    [0.577076292381737, "lognorm", [2.840507837788255, 0, 10810.683862254062]],
    [0.27234230800741915, "lognorm", [0.9060624216365498, 0, 69254.18734698475]],
    [0.11618815257568235, "lognorm", [1.4971495459456274, 0, 117536.41362290115]]
  ]
}
\end{lstlisting}

\subsubsection{Octets (length)}
\begin{lstlisting}[language=json]
{
  "sum": 244338999568258,
  "mix": [
    [2.5026115836964763e-05, "uniform", [0, 1]],
    [0.00030848682168057884, "uniform", [0, 2]],
    [0.02629300957158946, "lognorm", [1.4420675639316636, 0, 60.96961570186466]],
    [0.4321437483202767, "lognorm", [2.391763848296885, 0, 16292.10824934533]],
    [0.34651782224968797, "lognorm", [0.9126642203716411, 0, 95464.19089587945]],
    [0.19471190692092794, "lognorm", [2.443090503208728, 0, 99187.63657881475]]
  ]
}
\end{lstlisting}

\pagebreak

\subsubsection{Flows (size)}
\begin{lstlisting}[language=json]
{
  "sum": 2171291495,
  "mix": [
    [0.3562156807623126, "lognorm", [0.3907230406692569, 0, 97.44862683388862]],
    [0.24596152653055356, "lognorm", [0.8090562734237499, 0, 309.3964432303163]],
    [0.29936868220923935, "lognorm", [1.0767646679156089, 0, 2126.6544290721135]],
    [0.08064998077948095, "lognorm", [1.3082593320270717, 0, 25680.662146577415]],
    [0.015038557697568579, "lognorm", [1.351326892378382, 0, 416680.99610410555]],
    [0.00207471133617079, "lognorm", [1.082474779920202, 0, 5602171.637581343]],
    [0.0006826279721182512, "lognorm", [1.0191138296449576, 0, 53432349.30912044]],
    [8.232712543896472e-06, "lognorm", [1.1904677440253855, 0, 514468038.51135993]]
  ]
}
\end{lstlisting}

\subsubsection{Packets (size)}
\begin{lstlisting}[language=json]
{
  "sum": 264608398768,
  "mix": [
    [0.019110043221941177, "lognorm", [1.4651993205076452, 0, 768.2474092753541]],
    [0.19262668725066606, "lognorm", [2.294092315962418, 0, 150980.85680126044]],
    [0.361860135712086, "lognorm", [1.5607143019561114, 0, 5548432.350720047]],
    [0.23460483078887964, "lognorm", [0.9099214043924188, 0, 141799840.02481562]],
    [0.1917983030264225, "lognorm", [2.175778867451218, 0, 192372125.20865577]]
  ]
}
\end{lstlisting}

\subsubsection{Octets (size)}
\begin{lstlisting}[language=json]
{
  "sum": 244338999568258,
  "mix": [
    [0.0200921533485388, "lognorm", [1.9499336579584527, 0, 17318.938696766047]],
    [0.12687459714739766, "lognorm", [1.8942474190198826, 0, 1211417.6962635145]],
    [0.6660517733754548, "lognorm", [1.9205983111977711, 0, 82901773.48206052]],
    [0.1592969497877428, "lognorm", [0.5987334669415345, 0, 153610747.54291907]],
    [0.02768452634085962, "lognorm", [3.729862894501239, 0, 263013346.81005827]]
  ]
}
\end{lstlisting}

\pagebreak

\subsection{UDP-only}
\subsubsection{Flows (length)}
\begin{lstlisting}[language=json]
{
  "sum": 1737523167,
  "mix": [
    [0.5827340809061519, "uniform", [0, 1]],
    [0.18722586716889525, "uniform", [0, 2]],
    [0.09147052372747216, "uniform", [0, 3]],
    [0.05622351865268365, "uniform", [0, 4]],
    [0.029447542558586896, "lognorm", [0.48946755142877785, 0, 9.674436486545002]],
    [0.04022783768479337, "lognorm", [0.26026467622698696, 0, 4.8838363075201485]],
    [0.004722552787759103, "lognorm", [0.44942994700845956, 0, 29.41090858057708]],
    [0.005631263608965612, "lognorm", [0.9247173397792914, 0, 67.48121917707935]],
    [0.0018118271560744815, "lognorm", [1.0498087651873083, 0, 594.4921878194859]],
    [0.00032077017026583367, "lognorm", [1.8547729167446296, 0, 7892.503625889563]],
    [0.00015649931641790045, "lognorm", [1.1263636801013632, 0, 19824.000189566883]],
    [2.756716956778995e-05, "lognorm", [0.40128970829397104, 0, 90369.61088154344]],
    [1.4909236588009346e-07, "lognorm", [0.5733877781565124, 0, 12757029.472787634]]
  ]
}
\end{lstlisting}

\subsubsection{Packets (length)}
\begin{lstlisting}[language=json]
{
  "sum": 50730510162,
  "mix": [
    [0.015500700688304963, "uniform", [0, 1]],
    [0.009435060248085889, "uniform", [0, 2]],
    [0.0044931303724251555, "uniform", [0, 3]],
    [0.005412185554566266, "uniform", [0, 4]],
    [0.0016053021813832915, "uniform", [0, 5]],
    [0.004636517537471412, "uniform", [0, 6]],
    [0.002791076569731363, "uniform", [0, 7]],
    [0.015132565651191227, "lognorm", [0.7973580036468034, 0, 12.799911610401022]],
    [0.19438355109964933, "lognorm", [2.7929547227813876, 0, 8347.465832961101]],
    [0.16859560786890865, "lognorm", [0.5811494223637251, 0, 93439.76897592201]],
    [0.48811556105480514, "lognorm", [1.6416577840785012, 0, 212658.10830944445]],
    [0.08989874117347661, "lognorm", [0.663786121253159, 0, 16617548.048028754]]
  ]
}
\end{lstlisting}

\subsubsection{Octets (length)}
\begin{lstlisting}[language=json]
{
  "sum": 31098128445645,
  "mix": [
    [0.004294347339716763, "uniform", [0, 1]],
    [0.002828520769629071, "uniform", [0, 2]],
    [0.001066971856100983, "uniform", [0, 3]],
    [0.0017028778446197923, "uniform", [0, 4]],
    [0.0004732752524306332, "uniform", [0, 5]],
    [0.0011710470754648713, "uniform", [0, 6]],
    [0.0008916773469929783, "uniform", [0, 7]],
    [0.010034641229119045, "lognorm", [1.1076152185085193, 0, 22.255938063768316]],
    [0.23044127725909141, "lognorm", [2.2868576715155426, 0, 14556.169895186491]],
    [0.3148044616602549, "lognorm", [0.8072894441918375, 0, 97378.62560656141]],
    [0.28726231322202045, "lognorm", [1.1282822360447493, 0, 704887.045089726]],
    [0.1450285891445588, "lognorm", [0.550305719617275, 0, 19017262.81445133]]
  ]
}
\end{lstlisting}

\pagebreak

\subsubsection{Flows (size)}
\begin{lstlisting}[language=json]
{
  "sum": 1737523167,
  "mix": [
    [0.6368318183834879, "lognorm", [0.30605856160365386, 0, 119.194382163552]],
    [0.2028143298183967, "lognorm", [0.22637348632484486, 0, 319.1796675129598]],
    [0.1252950133909036, "lognorm", [0.5892942594845255, 0, 630.2029144235321]],
    [0.027699290695991776, "lognorm", [0.7595903510694968, 0, 3216.4358332108045]],
    [0.00593176349325996, "lognorm", [1.1418301435106715, 0, 29016.641311394553]],
    [0.0011888113464049218, "lognorm", [1.301077232574143, 0, 650546.3136647696]],
    [0.0001853823086593862, "lognorm", [1.708941299345426, 0, 12790970.28730909]],
    [5.349084480822565e-05, "lognorm", [1.2758486935847089, 0, 22546730.947471276]],
    [9.971808265531352e-08, "lognorm", [0.5244356726513165, 0, 21229598019.06612]]
  ]
}
\end{lstlisting}

\subsubsection{Packets (size)}
\begin{lstlisting}[language=json]
{
  "sum": 50730510162,
  "mix": [
    [0.010256204356373293, "lognorm", [0.17308480731567824, 0, 102.62679666811907]],
    [0.030844086441235714, "lognorm", [0.6432754020773255, 0, 267.6661032819417]],
    [0.02685706188133373, "lognorm", [1.4984158470482594, 0, 2965.85215348368]],
    [0.08580523855066874, "lognorm", [1.799761998854644, 0, 378347.651458613]],
    [0.18823579321779155, "lognorm", [1.0858777584070194, 0, 6754398.34678373]],
    [0.28805465555289284, "lognorm", [1.5399767728254337, 0, 42706686.55778615]],
    [0.22918391889821518, "lognorm", [1.4964054459366063, 0, 189231219.8306924]],
    [0.07888875545972893, "lognorm", [0.9014834641012421, 0, 1258964757.4953783]],
    [0.06187428564175922, "lognorm", [0.5087438892297369, 0, 27586008021.430702]]
  ]
}
\end{lstlisting}

\subsubsection{Octets (size)}
\begin{lstlisting}[language=json]
{
  "sum": 31098128445645,
  "mix": [
    [0.008743954302478988, "lognorm", [0.6556352655697212, 0, 233.0991244107341]],
    [0.009043458696784244, "lognorm", [1.2293570646603522, 0, 2081.1893487186962]],
    [0.01390867553910152, "lognorm", [1.7865980995011648, 0, 95940.98742907401]],
    [0.03790862213756647, "lognorm", [2.122469298512453, 0, 1384290.7533060464]],
    [0.1314371359145332, "lognorm", [1.3846650107480718, 0, 8243117.92583758]],
    [0.24317263385123908, "lognorm", [1.1587390305442005, 0, 112649756.51562566]],
    [0.27828358762231536, "lognorm", [1.1813069072741065, 0, 204868989.52410865]],
    [0.1428515782894115, "lognorm", [1.1064603547801617, 0, 1629156765.603726]],
    [0.13465035364656983, "lognorm", [0.49190322615873755, 0, 28150000653.69207]]
  ]
}
\end{lstlisting}

\end{document}